\documentclass[preprint,12pt]{elsarticle}

\usepackage{amsmath,amssymb,bm}
\usepackage{newtxtext,newtxmath}
\usepackage{graphicx}
\usepackage{booktabs}
\usepackage{siunitx}
\usepackage{url}
\usepackage{hyperref}
\hypersetup{hidelinks}

\journal{Chaos, Solitons \& Fractals}

\begin{document}
\begin{frontmatter}

\title{When will androids outnumber humans?}

\author[unimi]{Alessio Zaccone\corref{cor1}}
\ead{alessio.zaccone@unimi.it}
\ead[url]{https://orcid.org/0000-0002-6673-7043}
\author[exeterlsi,exeterphys]{Thomas Tunstall}
\ead{T.Tunstall@exeter.ac.uk}
\ead[url]{https://orcid.org/0000-0003-1945-9113}
\author[rockefeller]{Jesse H. Ausubel}
\ead{ausubel@rockefeller.edu}
\ead[url]{https://orcid.org/0000-0001-9568-9880}
\cortext[cor1]{Corresponding author}

\address[unimi]{Department of Physics ``A. Pontremoli'', University of Milan, via Celoria 16, 20133 Milan, Italy}
\address[exeterlsi]{Living Systems Institute, Faculty of Health and Life Sciences, University of Exeter, Exeter, United Kingdom}
\address[exeterphys]{Physics and Astronomy, Faculty of Environment, Science and Economy, University of Exeter, Exeter, United Kingdom}
\address[rockefeller]{Program for the Human Environment, The Rockefeller University, New York, NY 10065, USA}

\begin{abstract}
Humanoid robots are increasingly discussed not only as machines but also as a potential technological population. We develop a nonlinear population-dynamical framework to examine the factors governing their possible diffusion relative to humans, without assuming that large-scale adoption or numerical crossover is inevitable. Let $H(t)$ and $A(t)$ denote the human and android populations, respectively, and let $a=A/A_s$ be the normalized android stock. The human component is calibrated to the United Nations World Population Prospects 2024 median trajectory. The fitted android channel represents conventional, human-organized manufacturing and adoption and is calibrated to a conditional scenario anchored by approximately $10^4$ humanoids in 2025 and $10^9$ in 2050. It is described by $\dot a=(q_M+r_Ma)\exp(K_Ma^\gamma)$, where $q_M$ is a Direct source, $r_M$ a stock-dependent industrial-scaling coefficient, and $(K_M,\gamma)$ characterize the fitted nonlinear feedback. We separately introduce a hypothetical autonomous channel in which installed androids materially contribute to producing additional androids. This channel is not fitted to the 2025--2050 adoption trajectory. Under the minimal canonical TZ closure, with $r_A$ the baseline gross robot-mediated production coefficient, $\delta_A$ the effective retirement rate, and $K_R$ the autonomous feedback parameter, the dimensionless production--feedback number is $\mathcal{P}(a)=(r_A/\delta_A)\exp(K_Ra)$. The autonomous channel is locally self-amplifying if and only if $\mathcal{P}>1$, while numerical human--android parity, $A=H$, is a separate cumulative condition. Global uncertainty analysis shows that crossover dates remain strongly scenario-dependent, whereas the production--feedback threshold is structurally more robust. A deliberately extreme early-autonomy stress test can produce parity by 2050, but this is a conditional dynamical boundary rather than a technological forecast.
\end{abstract}


\end{frontmatter}

\section{Introduction}

Demographic models traditionally count biological populations. Rapid progress in humanoid robotics motivates a broader question: should a large installed stock of autonomous humanoid machines eventually be treated as a second population coexisting with humans? Chamie has explicitly proposed this demographic viewpoint, first in the context of ageing societies and later in terms of androids as a population category \cite{Chamie2017Robots,Chamie2022Androids,Chamie2025Androids}. In the most recent estimate, Chamie discusses a global stock of order $10^4$ androids in 2025 and a scenario of order $10^9$ by 2050 \cite{Chamie2025Androids}. Independently, Morgan Stanley Research projects more than one billion humanoid robots in use globally by 2050, with adoption remaining comparatively slow until the mid-2030s and accelerating thereafter \cite{MorganStanley2025Humanoids}. Corporate roadmaps likewise target physical labour and household deployment, although such statements are technological ambitions rather than demographic observations \cite{FigureMasterPlan}. For scale, the Japan Automobile Manufacturers Association reports that close to $1.70\times10^9$ motor vehicles (excluding motorcycles) were in use worldwide in 2023, showing that billion-unit stocks of complex manufactured machines are not unprecedented \cite{JAMA2025MotorVehicles}.

The possibility that advanced technological civilizations might eventually undergo a transition from biological to artificial, or post-biological, intelligence was already discussed in the Soviet SETI (Search for Extraterrestrial Intelligence) literature of the 1960s. Astrophysicist Iosif Shklovskii, in particular, considered the emergence of artificially intelligent beings as a potentially transformative stage---and even a possible crisis---in the evolution of a civilization. In his later work with Carl Sagan, the idea was discussed explicitly in the context of ``artificial intelligence and Galactic civilizations'', anticipating subsequent post-biological interpretations of the Fermi paradox \cite{Shklovskii1962,ShklovskiiSagan1966}.

Here we ask a narrower question: \emph{under what conditions can a manufactured technological population become numerically comparable to, or larger than, the biological population that created it?} By ``crossover'' we mean only numerical abundance, $A>H$, where $A(t)$ denotes the number of functioning humanoid robots and $H(t)$ the human population. It does not imply consciousness, political power, autonomy of goals, or displacement of humans.

Our modelling choice is motivated by two recent developments. First, the TZ population-growth equation has been shown to describe long-term human population regimes \cite{ZacconeTrachenko2026}. Second, the geometrical growth framework of Tunstall and Zaccone \cite{TunstallZaccone2026} identifies the canonical TZ law as the $\gamma=1$ Replicative member of a broader class and distinguishes Replicative growth from Direct production. This distinction is particularly natural here: human population growth is intrinsically replicative, whereas present-day humanoids are manufactured.

The extent to which future human and android populations will occupy and support the same productive network is presently unknown. We therefore make the absence of explicit cross-species network feedback a stated modelling assumption (Assumption A5), rather than building an unconstrained coupling into the calibrated baseline. A multi-species geometrical extension is developed in the Discussion to show how this assumption can be relaxed when suitable data become available.

The data-constrained model has three objectives: (i) calibrate the canonical TZ/Replicative growth law to the current United Nations demographic trajectory; (ii) calibrate and benchmark an effective externally organized android-adoption law containing Direct and stock-dependent industrial contributions; and (iii) derive, from the same TZ productivity-feedback logic, the conditions under which a distinct android-mediated Replicative channel becomes self-amplifying and can ultimately produce numerical crossover.

\section{Direct, Replicative, and TZ growth}
\label{sec:model_framework}

The geometrical framework introduced in Ref.~\cite{TunstallZaccone2026} distinguishes two basic growth mechanisms. In \emph{Replicative} growth, existing population elements contribute directly to the creation of new ones,
\begin{equation}
\frac{dX}{dt}
=
r_0 X\exp(\beta X^\gamma),
\label{eq:replicative_main}
\end{equation}
whereas in \emph{Direct} growth the population prefactor is absent,
\begin{equation}
\frac{dX}{dt}
=
q_0\exp(\beta X^\gamma).
\label{eq:direct_main}
\end{equation}
The parameters $r_0$ and $q_0$ set the corresponding rate scales, while $\gamma$ controls how the effective growth-supporting component space scales with population size. In the form written above, both growth classes are strictly positive for positive rate prefactors, so they cannot represent an exactly stationary or declining population. Zero or negative net growth can be incorporated by an affine rate transformation, as discussed in \cite{tunstall2026Geometric}. We retain the positive-growth form here for analytical tractability because the present calculations concern the timing of a rising android stock and are insensitive to this modification over the calibration and crossover windows considered. In particular, the slowing of the fitted android trajectory near 2050 should not be interpreted as a zero-growth plateau; the fitted instantaneous fractional growth rate remains positive there, as quantified below.

For $\gamma=1$, Eq.~(\ref{eq:replicative_main}) becomes
\begin{equation}
\frac{dX}{dt}
=
r_0 X\exp(\beta X),
\label{eq:replicative_gamma1}
\end{equation}
which is the canonical growth form of the Trachenko--Zaccone equation \cite{ZacconeTrachenko2026}. Recent work has shown that fitting the Replicative model to human populations over 1970--2023 recovers this canonical form \cite{TunstallZaccone2026}. We therefore use the following notation:
\begin{equation}
\dot h=r_Hh\exp(K_Hh).
\label{eq:humanTZ_bridge}
\end{equation}
For the human calibration considered here $K_H<0$, so the dimensionless biological feedback multiplier
\begin{equation}
\Phi_H(h)\equiv\exp(K_Hh)
\label{eq:human_feedback_multiplier}
\end{equation}
satisfies
\begin{equation}
0<\Phi_H(h)\leq1
\qquad (h\geq0).
\label{eq:human_feedback_bound}
\end{equation}
Thus population dependence can inhibit the baseline biological rate but does not amplify it above its zero-population reference value. The bound applies to the dimensionless feedback multiplier $\Phi_H$, not to the dimensional coefficient $r_H$ itself. A biological analogue of population-dependent collective regulation is quorum sensing, in which microbial behaviour changes when density-dependent signals cross a threshold \cite{WatersBassler2005}; the analogy is structural rather than mechanistic.

Androids require both growth classes. Present-day manufacturing is Direct in the sense that an existing humanoid need not reproduce for another to be built. The fitted conventional adoption law therefore combines a Direct source with a stock-dependent industrial-scaling contribution and allows a generalized exponent $\gamma\neq1$. By contrast, the hypothetical future channel in which installed androids materially contribute to producing additional androids is population-proportional and therefore Replicative. We model that new channel with the minimal canonical $\gamma=1$ TZ closure, not because the 2025--2050 baseline fit prefers $\gamma=1$---it does not---but because the autonomous term represents a different physical mechanism for which no production data yet identify a more general exponent. The canonical choice is therefore a parsimonious extrapolative closure, analogous to assuming that robot-mediated productivity scales with installed robot stock in the same first-order state variable used for the human Replicative channel.

The full pedagogical derivation connecting the TZ equation to Malthusian exponential growth, the weak-feedback Verhulst logistic and von Foerster/Riccati limits, and stretched- or compressed-exponential growth is given in the Appendix, Sec.~\ref{sec:s7_pedagogical_TZ}. Keeping that derivation separate allows the main text to proceed directly to the assumptions, calibration, and crossover analysis.

\section{Definitions, scope, and modelling assumptions}
\label{sec:assumptions}

Before constructing the dynamical equations, we state explicitly what is being counted and which physical and demographic effects are retained. The purpose of the model is not to predict social or political outcomes, but to identify the dynamical conditions under which the stock of functioning humanoid robots could become numerically comparable with the human population.

\paragraph{A1: Definition of the two populations}
$H(t)$ denotes the total world human population. $A(t)$ denotes the stock of functioning, deployed humanoid robots that are sufficiently general-purpose to be counted as individual physical units. Purely software agents, conventional industrial robot arms, autonomous vehicles, and non-humanoid specialized machines are not included in $A$. Each physical humanoid is counted once. This definition is deliberately operational because a demographic model requires a countable stock variable.

\paragraph{A2: Global mean-field description}
The first model is spatially homogeneous. We do not resolve countries, firms, factories, age classes, migration, supply chains, or spatial transport. Their aggregate effects enter only through effective rate parameters. A regional/network formulation is discussed later as a natural extension.

\paragraph{A3: Deterministic dynamics}
The equations describe expected global trajectories. Production shocks, wars, regulation, technological breakthroughs, and other stochastic events are not represented explicitly. Thus the trajectories should be interpreted as deterministic mean-field scenarios rather than sample paths of a stochastic process.

\paragraph{A4: Exogenous human baseline}
For calibration, humans are fitted independently of androids over 2025--2050. For the post-2050 crossover calculations the WPP2024 median human trajectory is used directly. Hence androids do not feed back onto human fertility or mortality in the baseline model. A cross-coupled extension is introduced in the Discussion.

\paragraph{A5: No explicit human--android network coupling in the calibrated baseline}
The calibrated baseline contains no explicit cross-species term in which the human stock directly amplifies android production or the android stock directly modifies human demographic rates. This is an identifiability assumption, not a claim that the two populations will be physically or economically independent. At present there is insufficient evidence to determine the sign, magnitude, or functional form of a combined human--android growth-supporting network. The future android-mediated channel is therefore modelled as being driven primarily by the installed android stock itself, while the human trajectory is supplied independently by WPP2024. If future cross-coupling is predominantly positive, omitting it will tend to delay the calculated crossover relative to a mutually supportive coupled model; adverse coupling could instead delay or prevent crossover. A multi-species geometrical extension that relaxes this assumption is developed in Sec.~\ref{subsec:multispecies}.

\paragraph{A6: Conditional android calibration}
The humanoid time series is not an observational census. The 2025 and 2050 values are scenario anchors from the cited demographic/market sources, while the intermediate 2030--2050 values are a rescaled timing proxy constructed from the published U.S. adoption scenario. Consequently the fitted android parameters are scenario parameters, not statistically inferred universal constants.

\paragraph{A7: Piecewise-constant effective parameters}
Within each calibration or scenario window, the effective coefficients governing
population growth, manufacturing, autonomous production, retirement, and nonlinear
feedback are treated as constant. This is the minimal closure adopted in the present
model. In reality, these quantities may evolve as economic costs, regulation,
resources, technological capabilities, and industrial organization change. The model
therefore identifies dynamical regimes and crossover conditions under specified
effective parameters rather than assuming that the underlying technological or
demographic environment remains unchanged indefinitely.

\paragraph{A8: Two physically distinct feedback channels}
The model distinguishes between two conceptually different mechanisms of
state-dependent feedback. The first describes the effective growth of the externally
organized manufacturing and adoption system, whereas the second describes the
hypothetical contribution of the installed android population to the production of
additional androids. Although both mechanisms are represented below using
state-dependent nonlinear rates, they correspond to physically distinct processes and
their feedback strengths are therefore not assumed to be identical.

\paragraph{A9: Canonical TZ closure for autonomous production}
When android-mediated production is introduced, we use the minimal canonical TZ assumption that the logarithm of gross productivity per active android varies linearly with the normalized android stock. This is a modelling closure for a new Replicative production mechanism, not a continuation of the fitted 2025--2050 baseline law. The baseline adoption channel is empirically generalized and gives $\gamma\neq1$, whereas no observations presently identify the stock dependence of robot-mediated productivity itself. We therefore adopt $\gamma=1$ as the lowest-complexity TZ closure, corresponding to the first-order assumption that autonomous productive capacity scales with the installed robot stock in the same canonical state variable used for the human Replicative channel in Ref.~\cite{ZacconeTrachenko2026}. The corresponding variables and coefficients are introduced explicitly below. More elaborate nonlinear state dependences should be tested when autonomous-production data become available.

\paragraph{A10: Numerical crossover is the only ``takeover'' criterion used mathematically}
The event studied quantitatively is
\begin{equation}
A(t_\star)=H(t_\star).
\label{eq:numerical_parity_definition}
\end{equation}
This is a statement about numbers only. It is not equivalent to economic dominance, political control, consciousness, autonomy of goals, or displacement of humans. We therefore use the terms \emph{parity} or \emph{numerical crossover} in the mathematical development.

These assumptions deliberately keep the first model minimal. Their advantage is that each additional physical effect---retirement, robot-mediated production, resource limitation, human--android coupling, or spatial fluxes---can be introduced as a separate term whose dynamical role is identifiable.

\section{Data and calibration}
\label{sec:data_calibration}

Throughout the paper, calendar time is measured in years. The human and android variables are normalized before fitting so that the nonlinear feedback parameters multiply dimensionless quantities. The numerical value of a feedback coefficient depends on the population scale used for normalization, whereas the combinations $K_H h$ and $K_M a^\gamma$ entering the exponential rates are invariant physical combinations within a specified normalization convention.

\subsection{Human population: normalization and TZ calibration}
\label{subsec:human_calibration}

We use the World Population Prospects 2024 (WPP2024) median projection of the United Nations Population Division \cite{UNWPP2024}. Over 2025--2050 the world population increases from approximately $8.23\times10^9$ to $9.66\times10^9$. We choose
\begin{equation}
H_s=10^{10}
\end{equation}
and define the dimensionless population
\begin{equation}
h(t)=\frac{H(t)}{H_s}.
\label{eq:h_definition}
\end{equation}
The human TZ growth law is
\begin{equation}
\frac{dh}{dt}=r_H h\exp(K_H h).
\label{eq:humanTZ}
\end{equation}
Because $h$ is dimensionless, $r_H$ has dimensions of inverse time and $K_H$ is dimensionless. The instantaneous per-capita human growth rate follows immediately by dividing Eq.~(\ref{eq:humanTZ}) by $h$:
\begin{equation}
g_H(h)
\equiv
\frac{1}{h}\frac{dh}{dt}
=
r_H\exp(K_H h).
\label{eq:human_percapita}
\end{equation}
Thus $K_H<0$ means that the per-capita growth rate decreases as population increases, while $K_H>0$ would represent a self-accelerating feedback. For the fitted human case the nonlinear factor $\Phi_H(h)=\exp(K_Hh)$ is therefore automatically bounded by unity, $0<\Phi_H\leq1$, consistent with the biological interpretation adopted in Sec.~\ref{sec:model_framework}. This bound concerns the dimensionless state-dependent multiplier, not the dimensional reference rate $r_H$.

Equation~(\ref{eq:humanTZ}) can also be integrated analytically. Separating variables gives
\begin{equation}
\frac{\exp(-K_H h)}{h}\,dh
=
r_H\,dt.
\label{eq:human_separation}
\end{equation}
Integrating from $(t_0,h_0)$ to $(t,h)$ yields
\begin{equation}
\int_{h_0}^{h}
\frac{\exp(-K_H u)}{u}\,du
=
r_H(t-t_0).
\label{eq:human_integral}
\end{equation}
Using $d\,\mathrm{Ei}(-K_Hu)/du=\exp(-K_Hu)/u$, one obtains the exact implicit solution
\begin{equation}
\mathrm{Ei}(-K_Hh)
-
\mathrm{Ei}(-K_Hh_0)
=
r_H(t-t_0),
\label{eq:human_Ei}
\end{equation}
where $\mathrm{Ei}$ is the exponential-integral function. In the limit $K_H\rightarrow0$, Eq.~(\ref{eq:humanTZ}) reduces to ordinary Malthusian growth,
\begin{equation}
h(t)=h_0\exp[r_H(t-t_0)].
\end{equation}

For the calibration we fix the initial condition to the WPP2024 value at $t_0=2025$,
\begin{equation}
h_0=\frac{H_{\rm WPP}(2025)}{H_s},
\end{equation}
and determine $r_H$ and $K_H$ by minimizing the unweighted normalized least-squares objective
\begin{equation}
\chi_H^2(r_H,K_H)
=
\frac{1}{N_H}
\sum_{i=1}^{N_H}
\left[
\frac{H_{\rm TZ}(t_i;r_H,K_H)-H_{\rm WPP}(t_i)}{H_s}
\right]^2,
\label{eq:human_objective}
\end{equation}
using the annual points from 2025 through 2050. This gives
\begin{equation}
r_H=0.2637\ {\rm yr}^{-1},
\qquad
K_H=-4.1325.
\label{eq:humanparams}
\end{equation}
The root-mean-square residual is approximately $5.2\times10^6$ persons and the maximum absolute residual over this interval is approximately $1.29\times10^7$ persons. These errors are small compared with a population scale of order $10^{10}$.

The fitted parameters have a useful rate interpretation. At the beginning of the interval, Eq.~(\ref{eq:human_percapita}) gives approximately
\begin{equation}
g_H(2025)\simeq8.8\times10^{-3}\ {\rm yr}^{-1},
\end{equation}
whereas by 2050 it has fallen to approximately
\begin{equation}
g_H(2050)\simeq4.9\times10^{-3}\ {\rm yr}^{-1}.
\end{equation}
Thus the negative $K_H$ compactly represents the demographic slowdown across the calibration window.

The constant-parameter solution of Eq.~(\ref{eq:humanTZ}) remains monotonically increasing, however, and therefore cannot reproduce the later WPP2024 population maximum and subsequent decline without allowing $r_H$ or $K_H$ to evolve. For this reason we use the TZ fit only as a compact description of 2025--2050. In all post-2050 crossover calculations the WPP2024 median trajectory $H_{\rm WPP}(t)$ itself is used as the human baseline.

\subsection{Android population: construction of the scenario timing proxy}
\label{subsec:android_anchors}

The android calibration is necessarily less direct because no census-quality global time series exists for general-purpose humanoids. We therefore distinguish source anchors from derived proxy points. Chamie gives the order-of-magnitude estimate $A(2025)=10^4$ for the global installed stock in 2025 \cite{Chamie2025Androids}. Morgan Stanley Research projects of order one billion humanoids globally by 2050 \cite{MorganStanley2025Humanoids}, so we use $A(2050)=10^9$ as the central high-adoption endpoint.

For the timing between these endpoints we use the detailed U.S. installed-base trajectory published by Morgan Stanley Research in its ``Humanoid 100'' report \cite{MorganStanleyHumanoid100}. The cumulative U.S. scenario is 0.04, 0.13, 0.50, 1.40, 3.80, 8.40, 16.20, 26.70, 38.50, 51.60 and 62.70 million units in 2030, 2032, \ldots, 2050, respectively. Because these are U.S. values rather than global forecasts, they are used only to define a dimensionless timing shape. Let $U_{\rm US}(t)$ denote this U.S. installed-base scenario and define
\begin{equation}
s(t)=\frac{U_{\rm US}(t)}{U_{\rm US}(2050)}.
\label{eq:timing_fraction}
\end{equation}
The global proxy points are
\begin{equation}
A_{\rm proxy}(t)=10^9 s(t),\qquad 2030\leq t\leq2050.
\label{eq:global_proxy}
\end{equation}
Together with $A(2025)=10^4$, Eq.~(\ref{eq:global_proxy}) supplies twelve timing points spanning five orders of magnitude. These intermediate values are not independent global observations or forecasts; they provide a reproducible way to transfer the published slow-then-accelerating timing shape to the global 2050 endpoint.

\subsection{Baseline android equation and endpoint-constrained calibration}
\label{subsec:android_calibration}

We normalize the deployed android stock by $A_s=10^9$ and write $a=A/A_s$. The conventional channel is intended to describe the net growth of the deployed stock generated by human-organized investment, factories, supply chains, demand and replacement activity. It is not yet a robot-reproduction law. We therefore use the generalized effective form
\begin{equation}
\frac{da}{dt}
=
(q_M+r_Ma)\exp(K_Ma^\gamma)
\label{eq:androidbaseline}
\end{equation}
with $q_M$ a Direct source, $r_Ma$ a stock-dependent industrial-scaling term, and $\exp(K_Ma^\gamma)$ an aggregate nonlinear modifier. Following the geometrical interpretation introduced by Tunstall and Zaccone \cite{TunstallZaccone2026}, the exponent $\gamma$ characterizes how the effective number of growth-supporting configurations or active interaction pathways scales with population size: $\gamma=1$ corresponds to the canonical TZ case, whereas $\gamma\neq1$ reflects a sublinear or superlinear expansion of the accessible growth-supporting component space as the population increases.

Both source anchors, $A_{\rm model}(2025)=10^4$ and $A_{\rm model}(2050)=10^9$, are enforced, while the intermediate timing-proxy points are fitted on the logarithmic population scale. The best fit gives
\begin{eqnarray}
q_M &=& 1.048\times10^{-5}\ {\rm yr}^{-1},\nonumber\\
r_M &=& 0.6459\ {\rm yr}^{-1},\nonumber\\
K_M &=& -1.8862,\nonumber\\
\gamma &=& 0.5944.
\label{eq:androidparams}
\end{eqnarray}
The log-RMSE over the twelve-point scenario trajectory is $0.0318$. Since $K_M<0$, the conventional channel slows as the deployed stock increases, but the fitted $\gamma<1$ spreads that inhibition over the trajectory rather than forcing an excessively sharp late flattening. At $a=1$, the nonlinear multiplier is $\exp(K_M)\simeq0.152$ and the instantaneous fractional baseline growth rate remains approximately $9.8\times10^{-2}$ yr$^{-1}$. Thus the 2050 endpoint is not a carrying-capacity plateau and, in particular, the positive-growth restriction discussed in Sec.~\ref{sec:model_framework} is not what prevents a plateau here: the fitted trajectory is still growing appreciably at the endpoint. An affine extension would become relevant only if one wished to represent an exactly stationary or declining android stock. More generally, an apparent limit over one epoch need not be permanent: bi-logistic models and logistic models with a time-varying carrying capacity describe successive growth pulses or technological changes that lift an earlier limit \cite{Meyer1994BiLogistic,MeyerAusubel1999CarryingCapacity}.

\begin{figure}[t]
\centering
\includegraphics[width=0.88\linewidth]{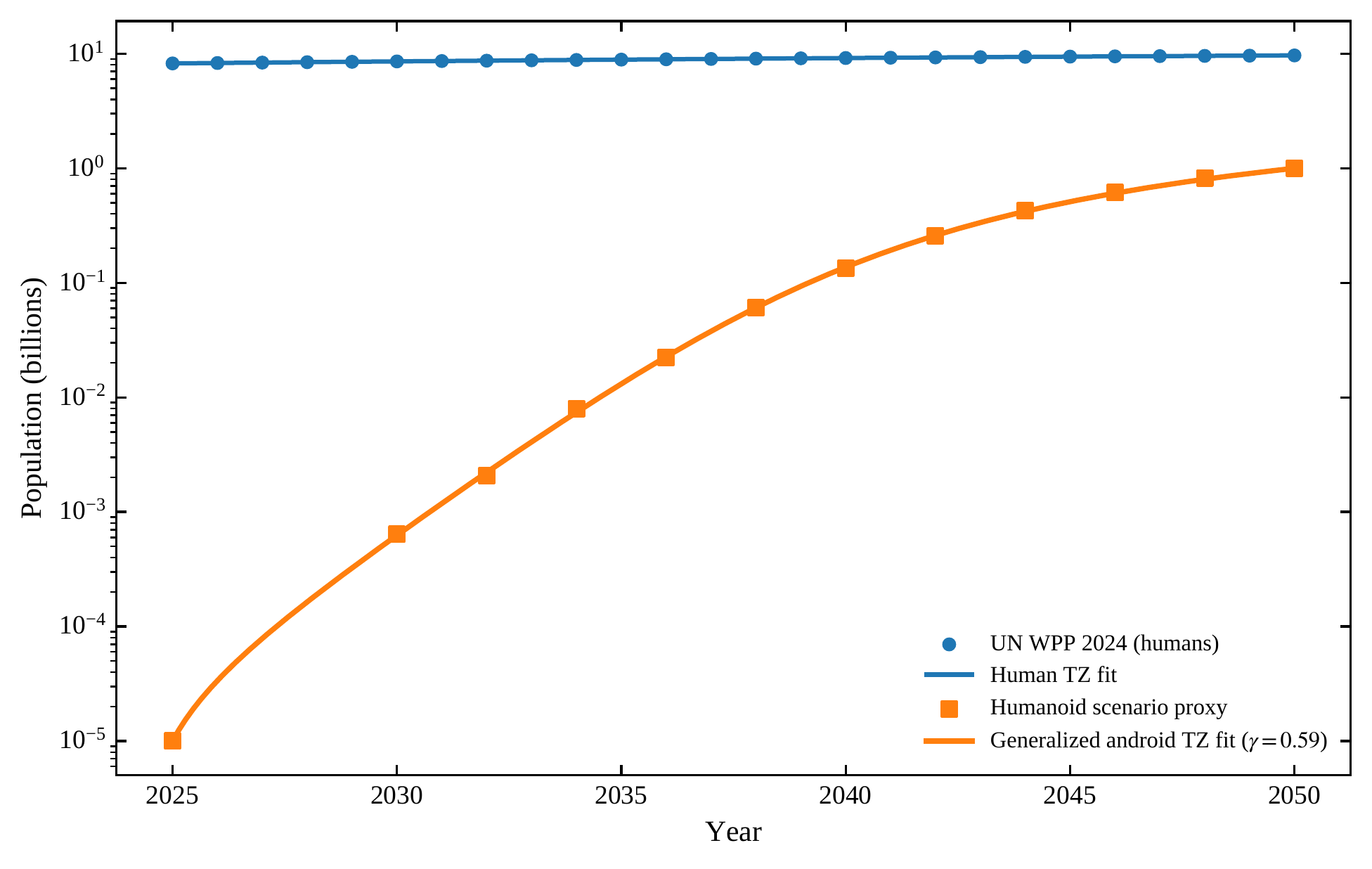}
\caption{Human and android calibration. Human points are the UN WPP2024 median trajectory and the line is the human TZ fit over 2025--2050. Android squares are the conditional timing proxy constructed from the 2025 global estimate, the detailed Morgan Stanley U.S. adoption timing, and the $10^9$ global 2050 endpoint. The android curve is the generalized endpoint-constrained fit, Eq.~(\ref{eq:androidbaseline}), with $\gamma\simeq0.59$. Its reduced logarithmic slope near 2050 reflects the adopted timing proxy and should not be interpreted as a finite carrying capacity near one billion units.}
\label{fig:calibration}
\end{figure}

\subsection{Numerical implementation}
\label{subsec:numerics}

All ordinary differential equations are integrated with an adaptive explicit Runge--Kutta method and tight relative/absolute tolerances; threshold parameters are obtained by one-dimensional root finding. The calculations were performed in Python/SciPy \cite{Virtanen2020SciPy}. For any specified parameter set, we integrate forward from the stated initial condition and evaluate the difference
\begin{equation}
D(t)=A(t)-H(t),
\label{eq:Ddefinition}
\end{equation}
and define the first numerical crossover as the earliest root $D(t_\star)=0$ for which $D$ changes from negative to non-negative. The source package supplied with the manuscript contains the calibration values and the data used to generate the figures.

\section{Benchmark against simpler adoption laws}
\label{sec:model_benchmark}

A nonlinear law should not be preferred merely because it has more parameters. We therefore benchmark the generalized baseline, Eq.~(\ref{eq:androidbaseline}), against endpoint-constrained exponential, logistic and Gompertz growth and against the canonical $\gamma=1$ version of the same TZ-inspired android law. Every model is forced through $A(2025)=10^4$ and $A(2050)=10^9$; only the intermediate timing shape is compared.

The standard benchmarks are
\begin{equation}
A(t)=A_0\exp[r(t-t_0)],
\end{equation}
\begin{equation}
\dot A=rA\left(1-\frac{A}{C}\right),
\end{equation}
and
\begin{equation}
\dot A=rA\ln\left(\frac{C}{A}\right),
\end{equation}
for exponential, logistic and Gompertz adoption, respectively. The canonical TZ-inspired android law is Eq.~(\ref{eq:androidbaseline}) with $\gamma=1$.

Because the scenario spans five orders of magnitude, goodness of fit is evaluated using residuals in $\ln A$. Table~\ref{tab:model_benchmark} reports log-RMSE and the small-sample Akaike information criterion AICc, using the number of remaining shape degrees of freedom after enforcing the two endpoint conditions \cite{BurnhamAnderson2002}. The generalized TZ law gives the smallest residual and the lowest AICc. The improvement relative to the canonical $\gamma=1$ form remains large even after penalizing the additional exponent: $\Delta$AICc $=21.8$.

\begin{table}[t]
\centering
\caption{Endpoint-constrained benchmark against the twelve-point humanoid timing proxy. The intermediate points inherit one Morgan Stanley scenario shape and are not independent observations.}
\label{tab:model_benchmark}
\begin{tabular}{lccc}
\toprule
Model & shape parameters & log-RMSE & $\Delta$AICc\\
\midrule
Exponential & 0 & 1.977 & 90.1\\
Logistic & 1 & 0.465 & 57.8\\
Gompertz & 1 & 0.194 & 36.8\\
Canonical TZ ($\gamma=1$) & 2 & 0.092 & 21.8\\
Generalized TZ & 3 & 0.032 & 0\\
\bottomrule
\end{tabular}
\end{table}

\begin{figure}[t]
\centering
\includegraphics[width=0.88\linewidth]{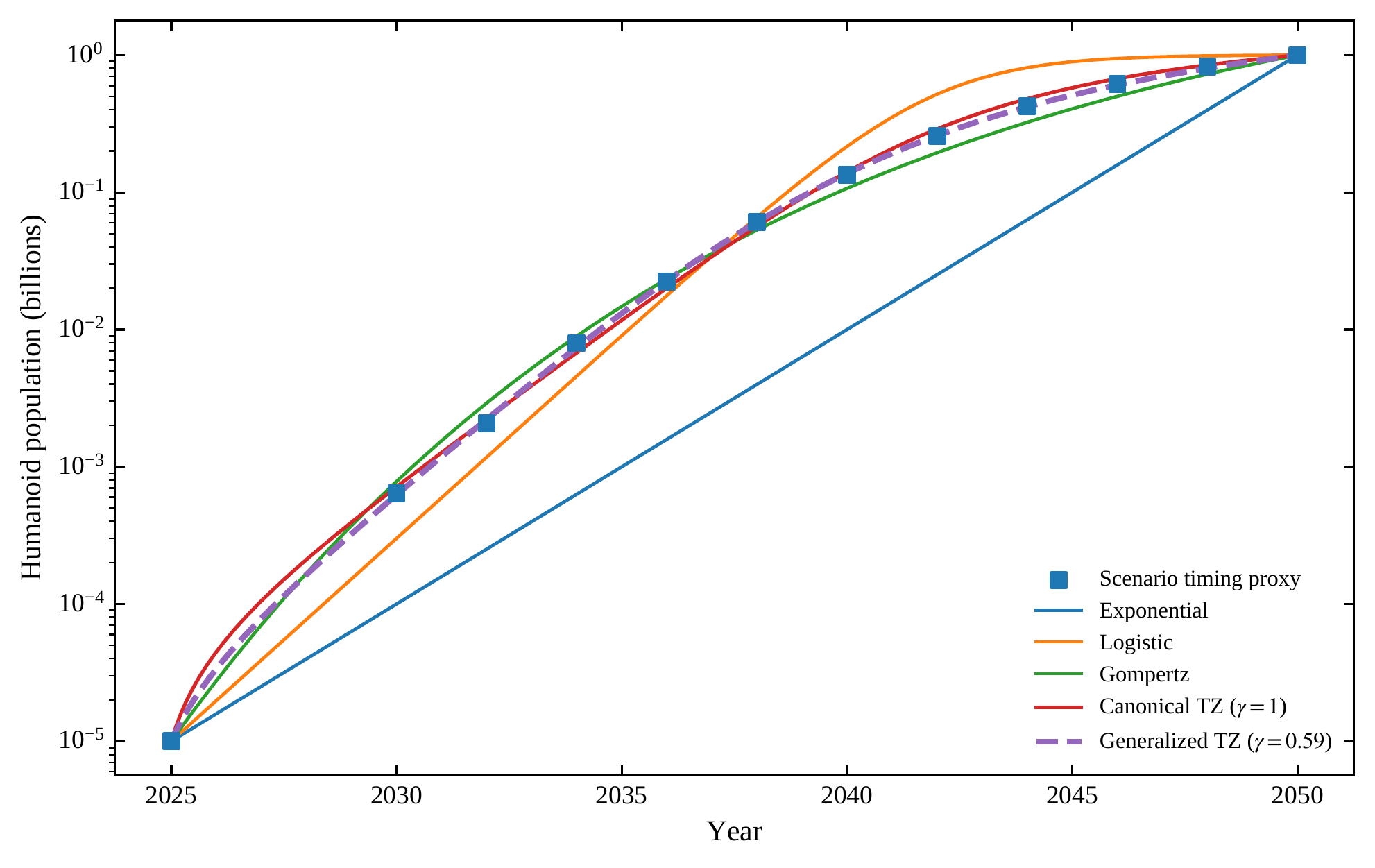}
\caption{Comparison of endpoint-constrained exponential, logistic, Gompertz, canonical TZ ($\gamma=1$), and generalized TZ adoption laws against the conditional humanoid timing proxy. The logarithmic vertical axis gives comparable weight to early and late parts of the five-decade population range.}
\label{fig:model_benchmark}
\end{figure}

This comparison does not establish that Eq.~(\ref{eq:androidbaseline}) is the unique law of future humanoid adoption: the target itself is forecast-derived. It does show that the extra exponent is not introduced simply to remove a visual feature in Fig.~\ref{fig:calibration}; it produces a substantial information-criterion improvement within the adopted timing scenario. Actual annual installed-base data will provide the decisive out-of-sample test.

\section{Nonlinear human--android dynamics}
\label{sec:nonlinear_dynamics}

The calibration above describes how the two populations evolve separately over the observed or scenario-constrained window. We now ask what additional conditions would allow the android population to gain share, become self-amplifying through its own productive contribution, and eventually reach numerical parity with humans. The analysis proceeds in stages. We first derive a model-independent composition criterion for crossover; we then separate the fitted conventional manufacturing feedback from a hypothetical robot-mediated Replicative channel; finally, we identify the fixed-point, self-amplification, production-autonomy and reduced crossover conditions associated with that new channel. This ordering is intended to distinguish clearly between kinematic statements that hold for any two populations and dynamical assumptions specific to autonomous android production.

\subsection{Composition dynamics and the exact crossover criterion}
\label{subsec:composition}

Before specifying any autonomous-production mechanism, we first isolate the purely kinematic requirement for one population to overtake the other. This provides a model-independent benchmark against which all later android-growth assumptions can be interpreted.

The total number of counted agents is
\begin{equation}
N(t)=A(t)+H(t).
\end{equation}
Rather than studying $N$ alone, the numerical competition between the two populations is described more directly by the android fraction
\begin{equation}
x(t)=\frac{A(t)}{A(t)+H(t)}=\frac{A}{N}.
\label{eq:android_fraction}
\end{equation}
The complementary human fraction is
\begin{equation}
1-x=\frac{H}{A+H}=\frac{H}{N}.
\end{equation}
Thus $x=0$ is a purely human population, $x=1$ a purely android population, and numerical parity $A=H$ corresponds exactly to
\begin{equation}
x=\frac{1}{2}.
\end{equation}

We now derive the equation for $x$ without assuming any particular functional form for $A(t)$ or $H(t)$. Differentiating Eq.~(\ref{eq:android_fraction}) by the quotient rule gives
\begin{eqnarray}
\dot x
&=&
\frac{\dot A(A+H)-A(\dot A+\dot H)}{(A+H)^2}
\nonumber\\
&=&
\frac{\dot A A+\dot A H-A\dot A-A\dot H}{(A+H)^2}
\nonumber\\
&=&
\frac{H\dot A-A\dot H}{(A+H)^2}.
\label{eq:x_derivation_1}
\end{eqnarray}
Introduce the instantaneous per-capita rates
\begin{equation}
g_A(t)=\frac{\dot A}{A},
\qquad
g_H(t)=\frac{\dot H}{H}.
\label{eq:percapita}
\end{equation}
Then $\dot A=A g_A$ and $\dot H=H g_H$, so the numerator of Eq.~(\ref{eq:x_derivation_1}) becomes
\begin{equation}
H\dot A-A\dot H
=
AH(g_A-g_H).
\end{equation}
Moreover,
\begin{equation}
\frac{AH}{(A+H)^2}
=
\frac{A}{A+H}\frac{H}{A+H}
=
x(1-x).
\label{eq:composition_identity}
\end{equation}
Substitution gives the exact composition equation
\begin{equation}
\dot x=x(1-x)\,[g_A(t)-g_H(t)].
\label{eq:replicator}
\end{equation}
No TZ assumption has entered Eq.~(\ref{eq:replicator}); it is a kinematic identity for any two positive populations. Since $x(1-x)>0$ for $0<x<1$, the sign of $\dot x$ is determined entirely by the relative per-capita rate:
\begin{equation}
\dot x>0
\quad\Longleftrightarrow\quad
g_A>g_H.
\label{eq:instantaneous_advantage}
\end{equation}
Hence androids gain population share whenever their instantaneous fractional growth rate exceeds the human fractional growth rate.

A still simpler representation uses the population ratio
\begin{equation}
R(t)=\frac{A(t)}{H(t)}.
\label{eq:R_definition}
\end{equation}
Taking a logarithm,
\begin{equation}
z(t)=\ln R(t)=\ln A(t)-\ln H(t).
\label{eq:logratio}
\end{equation}
Differentiating term by term gives
\begin{eqnarray}
\dot z
&=&
\frac{\dot A}{A}
-
\frac{\dot H}{H}
\nonumber\\
&=&
g_A(t)-g_H(t).
\label{eq:zdot}
\end{eqnarray}
Integration from $t_0$ to $t$ therefore yields
\begin{equation}
z(t)-z(t_0)
=
\int_{t_0}^{t}
[g_A(t')-g_H(t')]\,dt'.
\label{eq:z_integrated_step}
\end{equation}
Using $z(t_0)=\ln(A_0/H_0)$,
\begin{equation}
\ln\!\left[\frac{A(t)}{H(t)}\right]
=
\ln\!\left(\frac{A_0}{H_0}\right)
+
\int_{t_0}^{t}[g_A-g_H]dt'.
\label{eq:ratio_integrated}
\end{equation}
At parity, $A(t_\star)=H(t_\star)$ and the left-hand side vanishes. Therefore
\begin{equation}
\int_{t_0}^{t_\star}
[g_A(t)-g_H(t)]\,dt
=
\ln\!\left(\frac{H_0}{A_0}\right).
\label{eq:crossover}
\end{equation}
Equation~(\ref{eq:crossover}) is the exact numerical-crossover criterion. It says that the cumulative relative-growth advantage must compensate the initial logarithmic population gap. Using the 2025 scenario values gives
\begin{equation}
\ln\!\left[\frac{H(2025)}{A(2025)}\right]
\simeq
13.62,
\label{eq:initial_log_gap}
\end{equation}
illustrating the large integrated advantage required when starting from only $10^4$ humanoids.

\subsection{Two distinct feedback channels}
\label{subsec:full_android}

The exact crossover condition tells us how much relative growth advantage is required, but not what physical process could generate it. We therefore separate the android dynamics into two mechanisms: the conventional, human-organized manufacturing/adoption channel already calibrated to 2025--2050, and a hypothetical future channel in which installed androids materially contribute to producing additional androids. Keeping these channels distinct is essential because only the first is constrained by the baseline scenario data.

The calibrated conventional channel is
\begin{equation}
F_M^{\rm fit}(a)
=
(q_M+r_Ma)e^{K_Ma^\gamma}.
\label{eq:androidbaseline_recalled}
\end{equation}
Its two algebraic source contributions are
\begin{equation}
F_M^{\rm fit}(a)
=
q_Me^{K_Ma^\gamma}
+
r_Ma e^{K_Ma^\gamma}.
\label{eq:baseline_channels}
\end{equation}
The first is a Direct source and the second reflects scaling of the surrounding human-organized industrial system. Although the latter is proportional to $a$, it does not imply that androids manufacture androids. The entire function $F_M^{\rm fit}$ is inferred from the net deployed-stock scenario and should be interpreted as an effective conventional adoption channel.

The hypothetical robot-mediated channel is constructed differently. Its nonlinear feedback is not copied from the fitted $K_M$ term, and the use of the canonical $\gamma=1$ form is not a post-2050 switch in the baseline law. Rather, it is a separate closure for a new Replicative mechanism whose stock dependence is presently unmeasured. We use the canonical form as the lowest-complexity assumption: if each active android contributes to productive capacity through the robot-operated network, then the gross production term is proportional to the active android stock, while the logarithm of per-android productivity is taken to vary to first order with that same normalized stock. This is the robot analogue of the canonical TZ population logic used in Ref.~\cite{ZacconeTrachenko2026}. In that construction a population variable $y$ obeys
\begin{equation}
\dot y=r(y)y,
\label{eq:TZ_population_template}
\end{equation}
and the productivity per existing unit is assumed to have the exponential state dependence
\begin{equation}
r(y)=r_0e^{Ky}.
\label{eq:TZ_rate_template}
\end{equation}
Substitution gives the canonical TZ growth law
\begin{equation}
\dot y=r_0y e^{Ky}.
\label{eq:TZ_growth_template}
\end{equation}
Thus the feedback multiplies a population-proportional production term; $K>0$ increases productivity with population size and $K<0$ decreases it.

For android-mediated production, let $r_R(a)$ denote the gross rate at which one active android contributes, through the whole robot-operated industrial network, to the production capacity for new androids. The minimal TZ closure is
\begin{equation}
\ln\left[\frac{r_R(a)}{r_A}\right]
=
K_Ra,
\label{eq:robot_log_rate}
\end{equation}
or equivalently
\begin{equation}
r_R(a)=r_Ae^{K_Ra},
\qquad
r_A=\frac{1}{\tau_R}.
\label{eq:robot_rate}
\end{equation}
Here $r_A$ is the baseline gross production coefficient and $\tau_R$ is the corresponding characteristic time. Multiplying by the number of active androids gives the gross robot-mediated production term
\begin{equation}
B_R(a)
=
a r_R(a)
=
r_Aa e^{K_Ra}.
\label{eq:robot_birth_term}
\end{equation}
If androids retire or become obsolete with an effective constant hazard $\delta_A$, the isolated robot-mediated subsystem is therefore
\begin{equation}
\dot a
=
a\left[r_Ae^{K_Ra}-\delta_A\right].
\label{eq:autonomous_subsystem}
\end{equation}
Equation~(\ref{eq:autonomous_subsystem}) is the direct growth-sign analogue of the canonical TZ equation with a linear state variable in the exponent, supplemented only by a Malthus-type loss term. The demographic analogy can be carried one step further: $r_R(a)$ plays the role of a state-dependent gross technological fertility, whereas $\delta_A$ is an effective mortality or retirement hazard and $1/\delta_A$ the corresponding mean service lifetime. In demographic language, early technological generations may exhibit an analogue of high infant mortality and low child survival: early units can fail disproportionately often and have short service lives, while subsequent improvements in reliability, maintenance and design reduce the effective hazard and stabilize lifetimes. The constant-$\delta_A$ approximation used here should therefore be understood as a first-order cohort average rather than a claim of age-independent machine survival.

This derivation also explains why $K_R$ need not equal $K_M$. The pair $(K_M,\gamma)$ describes the aggregate state dependence of the externally organized market and manufacturing channel and is fitted to the 2025--2050 adoption scenario. By contrast, $K_R$ describes how the gross productivity \emph{per active android} changes once androids themselves materially contribute to the production chain. The mechanisms, observables and onset times are different. In the baseline calibration $r_A=0$ for this additional channel, so $K_R$ has no effect and cannot be inferred from the same data. Choosing the canonical exponent $K_Ra$ is therefore not a second fit of the same nonlinearity; it is the minimal TZ closure for a new physical channel.

A fully resolved stock-flow model would distinguish the gross externally organized production rate $G_M(a,t)$ from retirements:
\begin{equation}
\dot a
=
G_M(a,t)
+
r_Aa e^{K_Ra}
-
\delta_Aa.
\label{eq:android_full}
\end{equation}

The distinction anticipated in assumption A8 can now be stated explicitly in
mathematical terms. The nonlinear factor
$\exp(K_Ma^\gamma)$ appearing in the fitted baseline model,
Eq.~(\ref{eq:androidbaseline}), describes the effective state dependence of the
externally organized manufacturing and adoption channel. By contrast,
$\exp(K_Ra)$ describes the state-dependent productivity of the distinct
android-mediated production channel derived from the canonical TZ
population-growth construction. Accordingly, $K_M$ and $K_R$ quantify
different physical feedback mechanisms and are neither required nor expected
to have the same sign or magnitude. Moreover, whereas $K_M$ and $\gamma$ are
determined by fitting the baseline adoption trajectory, $K_R$ is a control
parameter describing a hypothetical future production regime and is not
constrained by the present calibration.

Present scenario data provide only the net installed-stock trajectory and do not separately identify $G_M$ and $\delta_A$. We therefore do not equate the fitted $F_M^{\rm fit}$ with a measured gross production flux. The quantitative combined-population calculations below use a reduced net closure, while Eq.~(\ref{eq:autonomous_subsystem}) is used for the physically interpretable self-amplification threshold.

\subsection{Autonomous subsystem: fixed points and stability}
\label{subsec:autonomous_subsystem}

For the robot-mediated channel considered on its own, the basic dynamical question is whether it decays, saturates, or becomes self-amplifying. The fixed points and their stability provide this classification independently of the human population and of the conventional manufacturing term.

Equation~(\ref{eq:autonomous_subsystem}) factorizes, so $a_0^\ast=0$ is always a fixed point. Differentiating the right-hand side gives
\begin{equation}
F'(0)=r_A-\delta_A,
\label{eq:zero_stability}
\end{equation}
so the zero state is stable if $r_A<\delta_A$ and unstable if $r_A>\delta_A$.

A non-zero fixed point satisfies
\begin{equation}
r_Ae^{K_Ra^\ast}=\delta_A.
\label{eq:nonzero_condition}
\end{equation}
Taking logarithms gives
\begin{equation}
a^\ast
=
\frac{1}{K_R}
\ln\left(\frac{\delta_A}{r_A}\right).
\label{eq:autonomous_fixedpoint}
\end{equation}
For stability, differentiate
\begin{equation}
F'(a)=r_Ae^{K_Ra}(1+K_Ra)-\delta_A.
\end{equation}
At the non-zero fixed point, $r_Ae^{K_Ra^\ast}=\delta_A$, hence
\begin{equation}
F'(a^\ast)=\delta_AK_Ra^\ast.
\label{eq:autonomous_stability}
\end{equation}
The sign of the non-zero eigenvalue is therefore exactly the sign of $K_R$.

For $K_R<0$ and $r_A>\delta_A$, the finite positive fixed point is stable and can be written
\begin{equation}
a^\ast
=
\frac{1}{|K_R|}
\ln\left(\frac{r_A}{\delta_A}\right).
\label{eq:stable_saturation}
\end{equation}
For $K_R<0$ and $r_A\leq\delta_A$, the zero state is stable and no positive fixed point exists. For $K_R=0$, Eq.~(\ref{eq:autonomous_subsystem}) reduces to $\dot a=(r_A-\delta_A)a$ with threshold $r_A=\delta_A$. For $K_R>0$ and $r_A<\delta_A$, the non-zero fixed point is unstable and acts as a critical-stock threshold: below it retirement dominates, whereas above it feedback makes the autonomous channel accelerate. Finally, for $K_R>0$ and $r_A\geq\delta_A$, the zero state is unstable and the minimal subsystem has no finite attracting fixed point.

\subsection{A dimensionless and falsifiable production--feedback threshold}
\label{subsec:critical_number}

The fixed-point analysis yields a dimensionless criterion that is directly interpretable and, in principle, measurable. This threshold separates retirement-dominated from locally self-amplifying autonomous production and is one of the main structural results of the model.

Measure time in units of the mean retirement time, $\theta=\delta_A(t-t_a)$, and define
\begin{equation}
\Pi=\frac{r_A}{\delta_A}.
\end{equation}
Equation~(\ref{eq:autonomous_subsystem}) becomes
\begin{equation}
\frac{da}{d\theta}
=
a\left[\Pi e^{K_Ra}-1\right].
\end{equation}
This identifies the state-dependent production--feedback number
\begin{equation}
\mathcal{P}(a)
=
\Pi e^{K_Ra}
=
\frac{r_A}{\delta_A}e^{K_Ra}.
\label{eq:P_number}
\end{equation}
Because $(1/a)da/d\theta=\mathcal{P}-1$, the transition is exact:
\begin{equation}
\begin{array}{lll}
\mathcal{P}<1 & : & \hbox{retirement dominated},\\
\mathcal{P}=1 & : & \hbox{threshold},\\
\mathcal{P}>1 & : & \hbox{self-amplifying}.
\end{array}
\label{eq:P_threshold}
\end{equation}
For $K_R>0$ and $\Pi<1$, the critical stock is $a_c=-\ln\Pi/K_R$; for $K_R<0$ and $\Pi>1$, the same condition gives the stable saturation stock.

\begin{figure}[t]
\centering
\includegraphics[width=0.80\linewidth]{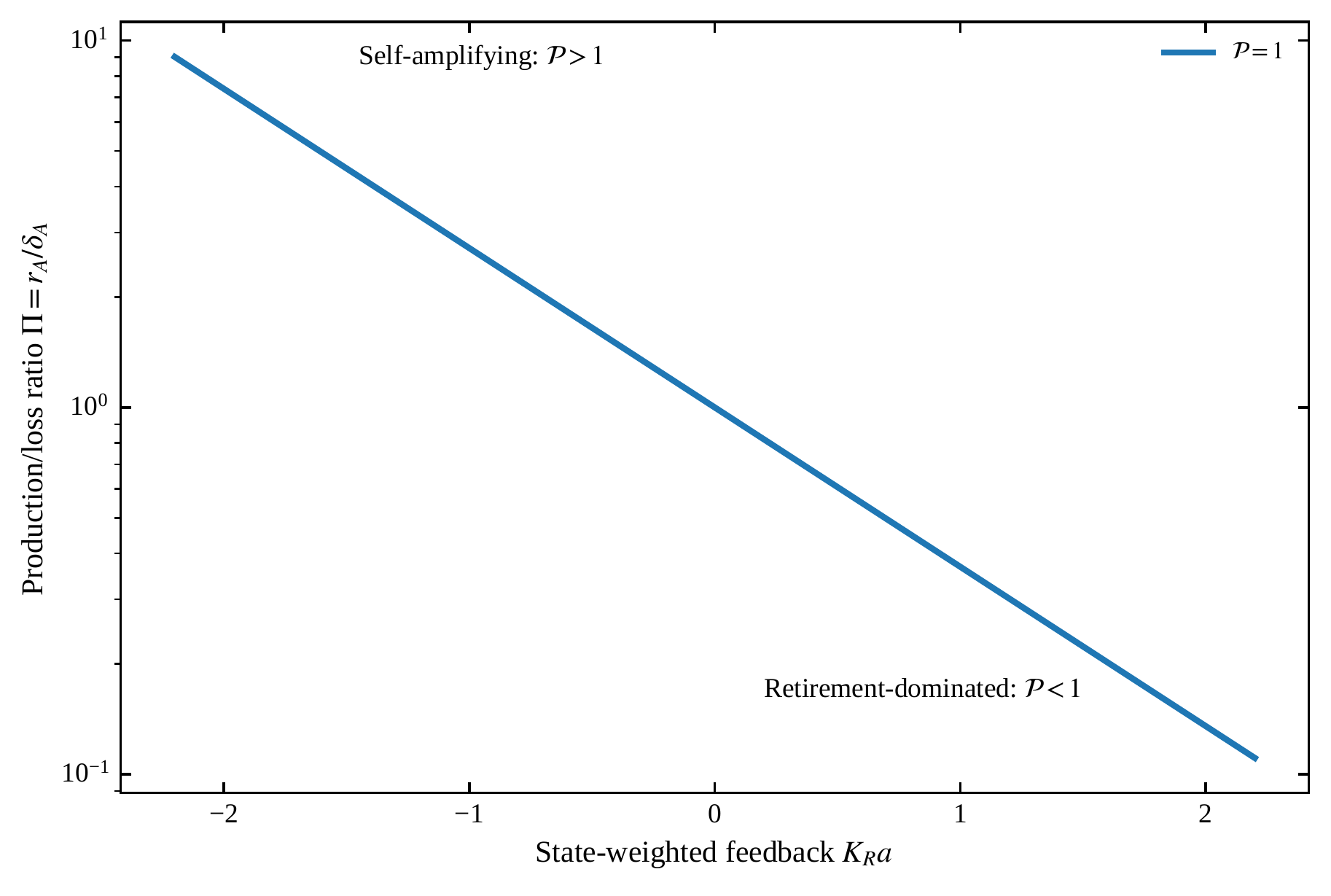}
\caption{Universal dimensionless transition of the canonical android-mediated TZ subsystem. The boundary between retirement-dominated and self-amplifying dynamics is $\mathcal{P}=\Pi e^{K_Ra}=1$, or $\Pi=e^{-K_Ra}$.}
\label{fig:dimensionless_threshold}
\end{figure}

\subsection{Production autonomy as a distinct threshold}
\label{subsec:autonomy_threshold}

Self-amplification of the isolated robot-mediated channel is not the same event as that channel becoming important relative to conventional manufacturing, and neither is identical to human--android parity. We therefore introduce a second diagnostic that asks when robot-mediated gross production becomes comparable with the fitted conventional-channel magnitude.

Numerical parity $A=H$ and production autonomy are conceptually distinct. As an operational diagnostic, compare the gross android-mediated term with the fitted conventional-channel magnitude:
\begin{equation}
\mathcal{R}_{\rm aut}^{\rm eff}(a)
=
\frac{r_Aa e^{K_Ra}}
{(q_M+r_Ma)e^{K_Ma^\gamma}}.
\label{eq:autonomy_ratio}
\end{equation}
Because the denominator is an effective net channel rather than a measured gross manufacturing flux, $\mathcal{R}_{\rm aut}^{\rm eff}$ is a scenario diagnostic rather than an exact industrial accounting identity. Its crossing nevertheless identifies the stock at which the new channel becomes comparable with the conventional one within the model. Setting $\mathcal{R}_{\rm aut}^{\rm eff}=1$ gives
\begin{equation}
\ln\left[
\frac{r_Aa_{\rm aut}}
{q_M+r_Ma_{\rm aut}}
\right]
+
K_Ra_{\rm aut}
-
K_Ma_{\rm aut}^\gamma
=0.
\label{eq:autonomy_exact}
\end{equation}
For $a\gg q_M/r_M$, this reduces to
\begin{equation}
\ln\left(\frac{r_A}{r_M}\right)
+
K_Ra_{\rm aut}
-
K_Ma_{\rm aut}^\gamma
\simeq0.
\label{eq:autonomy_approx}
\end{equation}
which is generally solved numerically. Only in the special canonical case $\gamma=1$ does it reduce to the elementary expression $a_{\rm aut}\simeq\ln(r_M/r_A)/(K_R-K_M)$.

\subsection{Reduced net model used for crossover scenarios}
\label{subsec:reduced_model}

The previous subsections identify physically interpretable thresholds, but the available scenario data do not separately determine gross robot-mediated production and retirement. To connect those thresholds to numerical crossover calculations without introducing unidentifiable stock-flow parameters, we therefore use a reduced net closure for the scenario analysis that follows.

Because gross robot-mediated production and retirement cannot yet be measured separately, the combined human--android scenario calculations use the minimal reduced equation
\begin{equation}
\dot a
=
(q_M+r_Ma)e^{K_Ma^\gamma}
+
\mu a e^{K_Ra}.
\label{eq:extended}
\end{equation}
Here the first term is the calibrated effective conventional channel and the second is a net additional android-mediated contribution. If one starts from the isolated gross/loss expression in Eq.~(\ref{eq:autonomous_subsystem}), its exact factorization is
\begin{equation}
r_Aa e^{K_Ra}-\delta_Aa
=
\left[r_A-\delta_Ae^{-K_Ra}\right]a e^{K_Ra},
\end{equation}
so the exact effective coefficient would be
\begin{equation}
\mu_{\rm eff}(a)=r_A-\delta_Ae^{-K_Ra}.
\label{eq:mueff}
\end{equation}
For $K_R=0$ this reduces exactly to
\begin{equation}
\mu=r_A-\delta_A.
\label{eq:mu_neutral}
\end{equation}
For $K_R\neq0$, a constant $\mu$ is a deliberate coarse-graining over the stock range of interest. The physically sharper statement is therefore the threshold $\mathcal{P}=1$ in the isolated autonomous subsystem, while Eq.~(\ref{eq:extended}) is used only to map conditional crossover scenarios.

\section{Results}
\label{sec:results}

\subsection{Baseline continuation and interpretation}
\label{subsec:baseline_results}

The generalized android equation is calibrated only through 2050. To understand what its fitted slowdown implies if the same effective parameters were to persist, we first continue Eq.~(\ref{eq:androidbaseline}) beyond the calibration window. This is an extrapolation, not an additional fit. Starting from $a(2050)=1$, the continuation gives
\begin{equation}
A_{\rm base}(2100)
\simeq
3.96\times10^9,
\end{equation}
whereas WPP2024 gives $H_{\rm WPP}(2100)\simeq1.019\times10^{10}$. Hence
\begin{equation}
\frac{A_{\rm base}(2100)}{H_{\rm WPP}(2100)}
\simeq
0.388,
\end{equation}
and numerical parity is not reached in the baseline continuation. Although this baseline is not a constant-rate exponential, its increase from $10^9$ units in 2050 to $3.96\times10^9$ in 2100 corresponds to an endpoint-equivalent average CAGR of approximately $2.79\%$ yr$^{-1}$.

The generalized fit also changes the interpretation of the apparent 2050 flattening. At $a=1$,
\begin{equation}
e^{K_Ma^\gamma}
=
e^{-1.8862}
\simeq
0.152,
\end{equation}
but the larger fitted $r_M$ leaves an instantaneous fractional baseline growth rate of approximately $0.098$ yr$^{-1}$. The trajectory is therefore still increasing appreciably at 2050; the curvature in Fig.~\ref{fig:calibration} should not be read as saturation at one billion units.

We next activate the reduced additional channel in Eq.~(\ref{eq:extended}) at 2050. For the illustrative value $\mu=0.04$ yr$^{-1}$, negative feedback $K_R=-0.05$ produces parity only at the very end of the horizon (about 2100), neutral feedback $K_R=0$ produces parity at approximately 2090, and positive feedback $K_R=+0.05$ advances parity to approximately 2083. These trajectories isolate the role of the autonomous state feedback at a fixed net coefficient. Because the instantaneous growth rate in these nonlinear trajectories varies with the state $a(t)$, none of them possesses a unique constant compound growth rate; any quoted CAGR would only be an endpoint-equivalent average and would obscure the feedback that the model is designed to resolve.

\begin{figure}[t]
\centering
\includegraphics[width=0.88\linewidth]{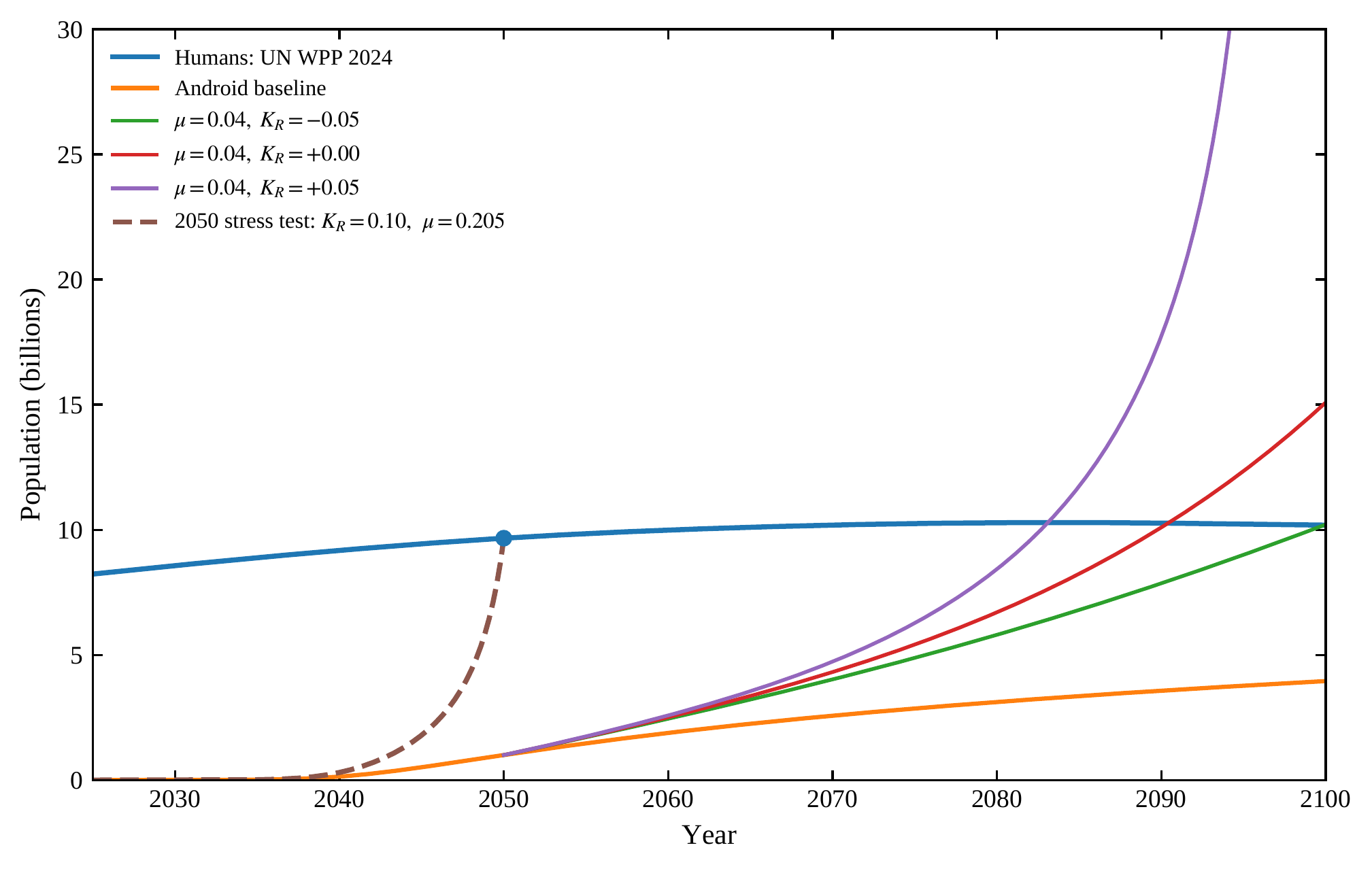}
\caption{Conditional human--android trajectories using the generalized conventional channel. The human curve follows WPP2024 and the android baseline is the continuation of Eq.~(\ref{eq:androidbaseline}). Three post-2050 curves use Eq.~(\ref{eq:extended}) with $\mu=0.04$ yr$^{-1}$ and different $K_R$. The dashed early-autonomy curve is a deliberately extreme stress test: the additional channel is activated in 2035 with $K_R=0.10$ and $\mu=0.205$ yr$^{-1}$, which is the threshold required in this calibration to obtain $A=H$ in 2050.}
\label{fig:scenarios}
\end{figure}

\subsection{A deliberately extreme 2050 crossover stress test}
\label{subsec:2050_stress}

The baseline calibration contains approximately $10^9$ humanoids in 2050, compared with a WPP2024 human population of approximately $9.66\times10^9$ \cite{UNWPP2024,MorganStanley2025Humanoids}. Therefore a numerical android majority in 2050 is not a consequence of the baseline scenario. It requires a much more aggressive trajectory. We quantify the departure from the baseline in two steps.

\paragraph{Step 1: model-independent exponential benchmark.}
Suppose, only for this benchmark, that the android stock obeyed constant-rate exponential growth
\begin{equation}
\dot A=rA,
\end{equation}
where in this paragraph $r$ is a constant exponential rate (not the parameter $r_A$ of the life-cycle model). The solution is
\begin{equation}
A(t)=A_0e^{r(t-t_0)}.
\end{equation}
Demanding $A(2050)=H(2050)$ gives
\begin{equation}
H(2050)
=
A(2025)e^{25r},
\end{equation}
so
\begin{eqnarray}
r_{2050}
&=&
\frac{1}{25}
\ln\!\left[
\frac{H(2050)}{A(2025)}
\right]
\nonumber\\
&=&
\frac{1}{25}
\ln\!\left(
\frac{9.664\times10^9}{10^4}
\right)
\nonumber\\
&\simeq&
0.551\ {\rm yr}^{-1}.
\label{eq:2050_benchmark}
\end{eqnarray}
The equivalent discrete compound annual growth rate is
\begin{equation}
{\rm CAGR}
=
e^{r_{2050}}-1
\simeq0.735,
\end{equation}
Thus the constant-exponential benchmark corresponds to a continuous rate of $55.1\%$ yr$^{-1}$ and an equivalent discrete CAGR of approximately $73.5\%$ per year for 25 consecutive years. This benchmark shows immediately that 2050 parity belongs to a very high-growth tail of possible scenarios.

\paragraph{Step 2: nonlinear early-autonomy stress test.}
We next retain the calibrated baseline dynamics until an onset time $t_a$ and activate the additional channel only after that date. Thus
\begin{equation}
\dot a
=
(q_M+r_Ma)e^{K_Ma^\gamma},
\qquad t<t_a,
\end{equation}
and
\begin{equation}
\dot a
=
(q_M+r_Ma)e^{K_Ma^\gamma}
+
\mu a e^{K_Ra},
\qquad t\ge t_a.
\label{eq:piecewise_stress}
\end{equation}
Continuity requires the initial value for the second stage to equal the first-stage solution,
\begin{equation}
a(t_a^+)=a(t_a^-)=a_{\rm base}(t_a).
\label{eq:continuity_onset}
\end{equation}
For reference, the calibrated baseline gives approximately $6.26\times10^5$ androids in 2030, $1.31\times10^7$ in 2035, and $1.37\times10^8$ in 2040. The later the autonomous channel starts, the larger its required coefficient must therefore be to close the remaining gap by 2050.

For fixed $(t_a,K_R)$, the threshold $\mu_c$ is defined by the terminal-value condition
\begin{equation}
F(\mu_c;t_a,K_R)
\equiv
A(2050;t_a,\mu_c,K_R)-H(2050)
=0.
\label{eq:2050_threshold_function}
\end{equation}
Because the final android stock is monotone in $\mu$ over the explored parameter range, the root is unique and can be obtained by scalar root finding. Taking $t_a=2035$ gives
\begin{equation}
\mu_c=0.269\ {\rm yr}^{-1}
\qquad(K_R=0),
\end{equation}
while positive feedback $K_R=0.10$ lowers the threshold to
\begin{equation}
\mu_c=0.205\ {\rm yr}^{-1}
\qquad(K_R=0.10).
\label{eq:2050_stress}
\end{equation}
The dashed trajectory in Fig.~\ref{fig:scenarios} uses the latter pair and reaches $A=H$ at approximately 2050 by construction.

Table~\ref{tab:2050stress} shows the sensitivity to onset time and feedback. Two trends are clear. First, delaying onset increases $\mu_c$ because less time remains for compounding. Second, increasing $K_R$ lowers $\mu_c$ because each increment in stock amplifies the autonomous rate more strongly.

\begin{table}[t]
\centering
\caption{Net autonomous coefficient $\mu_c$ (yr$^{-1}$) required for numerical parity exactly in 2050, as a function of onset year and nonlinear feedback $K_R$. These values define stress-test thresholds, not forecasts.}
\label{tab:2050stress}
\begin{tabular}{ccccc}
\toprule
Onset year & $K_R=-0.05$ & $K_R=0$ & $K_R=+0.05$ & $K_R=+0.10$\\
\midrule
2030 & 0.273 & 0.237 & 0.207 & 0.183\\
2035 & 0.313 & 0.269 & 0.233 & 0.205\\
2040 & 0.389 & 0.331 & 0.285 & 0.248\\
\bottomrule
\end{tabular}
\end{table}

\subsection{Analytical benchmark for neutral post-2050 feedback}
\label{subsec:2100_benchmark}

The 2100 numerical threshold can also be understood without solving the full ODE. Suppose that after 2050 the additional channel is neutral, $K_R=0$, and dominates the incremental growth. Neglect the calibrated baseline term only for this estimate. Then
\begin{equation}
\dot A\simeq\mu A.
\end{equation}
Separating variables gives
\begin{equation}
\frac{dA}{A}=\mu\,dt,
\end{equation}
and integrating from $t_0$ to $t_1$ gives
\begin{equation}
\ln\!\left[\frac{A(t_1)}{A(t_0)}\right]
=
\mu(t_1-t_0).
\end{equation}
Requiring parity $A(t_1)=H(t_1)$ therefore gives
\begin{equation}
\mu_c^{(0)}
=
\frac{1}{t_1-t_0}
\ln\!\left[\frac{H(t_1)}{A(t_0)}\right].
\label{eq:analytic_mucrit}
\end{equation}
With $t_0=2050$, $A(t_0)=10^9$, $t_1=2100$, and $H(t_1)=10.1866\times10^9$,
\begin{eqnarray}
\mu_c^{(0)}
&=&
\frac{1}{50}\ln(10.1866)
\nonumber\\
&\simeq&
0.0464\ {\rm yr}^{-1}.
\label{eq:analytic_mu_value}
\end{eqnarray}
This corresponds to a continuous exponential rate of about $4.64\%$ yr$^{-1}$, equivalent to a discrete CAGR of approximately $4.75\%$ yr$^{-1}$, in the simplified channel. The full calibrated model gives the lower threshold
\begin{equation}
\mu_c\simeq0.0300\ {\rm yr}^{-1},
\end{equation}
because the pre-existing baseline term in Eq.~(\ref{eq:extended}) continues to add androids throughout 2050--2100. The difference between $0.0464$ and $0.0300$ yr$^{-1}$ therefore measures the contribution of the baseline manufacturing/adoption channel within this calibration.

\subsection{Construction of the crossover phase diagram}
\label{subsec:phase_diagram}

\begin{figure}[t]
\centering
\includegraphics[width=0.86\linewidth]{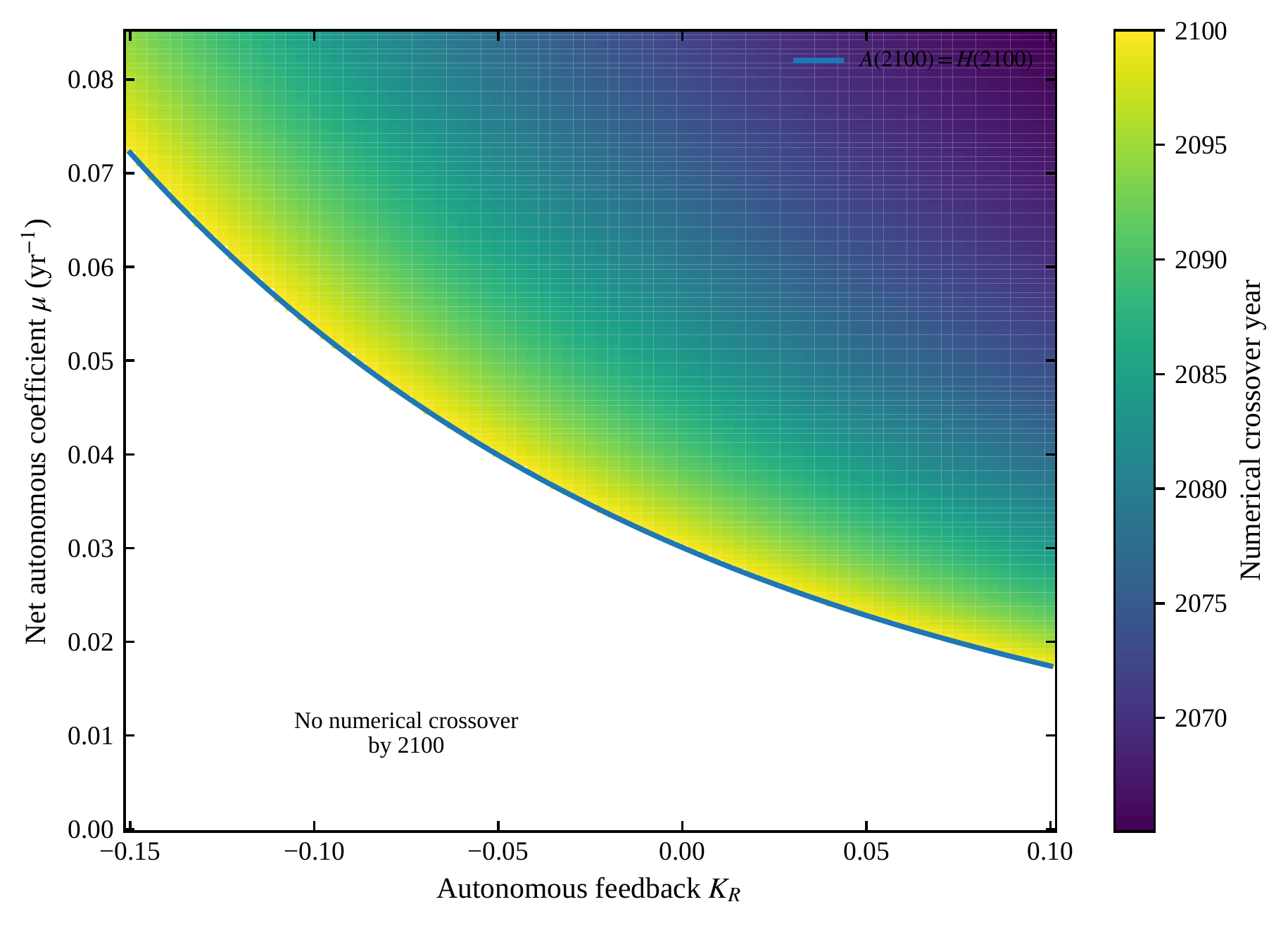}
\caption{Crossover phase diagram for Eq.~(\ref{eq:extended}). The color field gives the first numerical-crossover year $A=H$; blank regions do not reach parity by 2100. The line denotes the boundary $A(2100)=H(2100)$. Positive autonomous feedback lowers the required net autonomous-production coefficient, whereas negative feedback raises it.}
\label{fig:phase}
\end{figure}

The phase diagram in Fig.~\ref{fig:phase} is constructed as follows. We set
\begin{equation}
a(2050)=1
\end{equation}
from the calibrated scenario endpoint and choose a pair $(K_R,\mu)$. We then integrate Eq.~(\ref{eq:extended}) forward to 2100 while reading the human population from WPP2024. At every time step we evaluate
\begin{equation}
D(t;K_R,\mu)=A(t;K_R,\mu)-H_{\rm WPP}(t).
\label{eq:phase_D}
\end{equation}
If $D$ first changes sign from negative to non-negative, that first root is recorded as the crossover year. If no such root occurs before 2100, the parameter pair is classified as ``no crossover by 2100.''

For a fixed $K_R$, the boundary of the crossover region is found by solving
\begin{equation}
A(2100;K_R,\mu_c)=H_{\rm WPP}(2100).
\label{eq:phase_boundary_condition}
\end{equation}
Within the explored range the android solution is monotone in $\mu$, so this scalar equation has a unique threshold. At neutral feedback,
\begin{equation}
\mu_c(K_R=0)\simeq0.0300\ {\rm yr}^{-1}.
\label{eq:mucrit_2100}
\end{equation}
Representative values are given in Table~\ref{tab:thresholds}. The monotonic trend is physically intuitive: positive $K_R$ amplifies the autonomous channel as $a$ grows and therefore reduces the required base coefficient $\mu$; negative $K_R$ progressively suppresses it and raises the threshold.

\begin{table}[t]
\centering
\caption{Critical net autonomous-production coefficient for numerical parity by 2100.}
\label{tab:thresholds}
\begin{tabular}{cc}
\toprule
$K_R$ & $\mu_c$ (yr$^{-1}$)\\
\midrule
-0.15 & 0.072\\
-0.10 & 0.053\\
-0.05 & 0.040\\
0     & 0.030\\
+0.05 & 0.023\\
+0.10 & 0.017\\
\bottomrule
\end{tabular}
\end{table}

For $K_R=0$, the reduced coefficient has the exact life-cycle interpretation $\mu=r_A-\delta_A$. If the mean android service lifetime were ten years and retirement were approximately memoryless, then
\begin{equation}
\delta_A\simeq\frac{1}{10}\ {\rm yr}^{-1}=0.1\ {\rm yr}^{-1}.
\end{equation}
The neutral-feedback 2100 threshold would then correspond to a gross autonomous-production coefficient
\begin{eqnarray}
r_A
&=&
\mu_c+\delta_A
\nonumber\\
&\simeq&
0.0300+0.1000
\nonumber\\
&\simeq&
0.130\ {\rm yr}^{-1}.
\end{eqnarray}
This conversion is illustrative only: actual robot retirement is unlikely to be exactly memoryless and will eventually require an age-structured survival model.

\subsection{Robustness to parameter uncertainty}
\label{subsec:uncertainty}

The quantitative crossover thresholds necessarily depend on quantities that are
presently poorly constrained, particularly the installed android population, future
deployment rates, the onset of android-mediated production, service lifetime, and
nonlinear production feedback. We therefore tested the robustness of the results by
means of a global uncertainty and sensitivity analysis. The complete methodology,
parameter ranges, sampling procedure, threshold distributions, and global sensitivity
analysis are reported in the Appendix, Secs.~\ref{sec:s1}--\ref{sec:s5}.

Briefly, we performed a 4096-point Latin-hypercube exploration of the uncertain parameters, varying the 2025 android population, the 2050 population endpoint, the onset year $t_a$ of android-mediated production, the mean service lifetime $L=1/\delta_A$, and the nonlinear feedback parameter $K_R$. The adopted ranges are deliberately broad epistemic envelopes rather than statistical confidence intervals.

Across this uncertainty ensemble, the critical gross android-mediated production coefficient required for numerical parity by 2100 has median
\begin{equation}
r_{A,c}\simeq 0.178~{\rm yr}^{-1},
\label{eq:uncertainty_median}
\end{equation}
with a 95\% design interval approximately
\begin{equation}
0.095 < r_{A,c} < 0.340~{\rm yr}^{-1}.
\label{eq:uncertainty_interval}
\end{equation}
The corresponding dimensionless production-to-loss ratio $\Pi_c=r_{A,c}/\delta_A$ has a median value of approximately $2.08$. Global rank sensitivity analysis shows that the threshold is controlled primarily by the mean service lifetime, the onset of android-mediated production, and the nonlinear feedback parameter $K_R$, whereas uncertainty in the absolute 2025 android population has a substantially weaker influence.

These results show that the numerical crossover year is intrinsically uncertain at present, whereas the existence of the production--feedback threshold derived in Sec.~\ref{subsec:critical_number} is structurally robust. Full parameter ranges, sampling methodology, uncertainty distributions, and
rank-sensitivity coefficients are provided in the Appendix
(Secs.~\ref{sec:s1}--\ref{sec:s5}, Figs.~\ref{fig:uncertainty_distribution}--\ref{fig:rank_sensitivity} and Table~\ref{tab:sensitivity}).

\subsection{Four dynamical regimes}
\label{subsec:four_regimes}

\begin{figure}[t]
\centering
\includegraphics[width=0.94\linewidth]{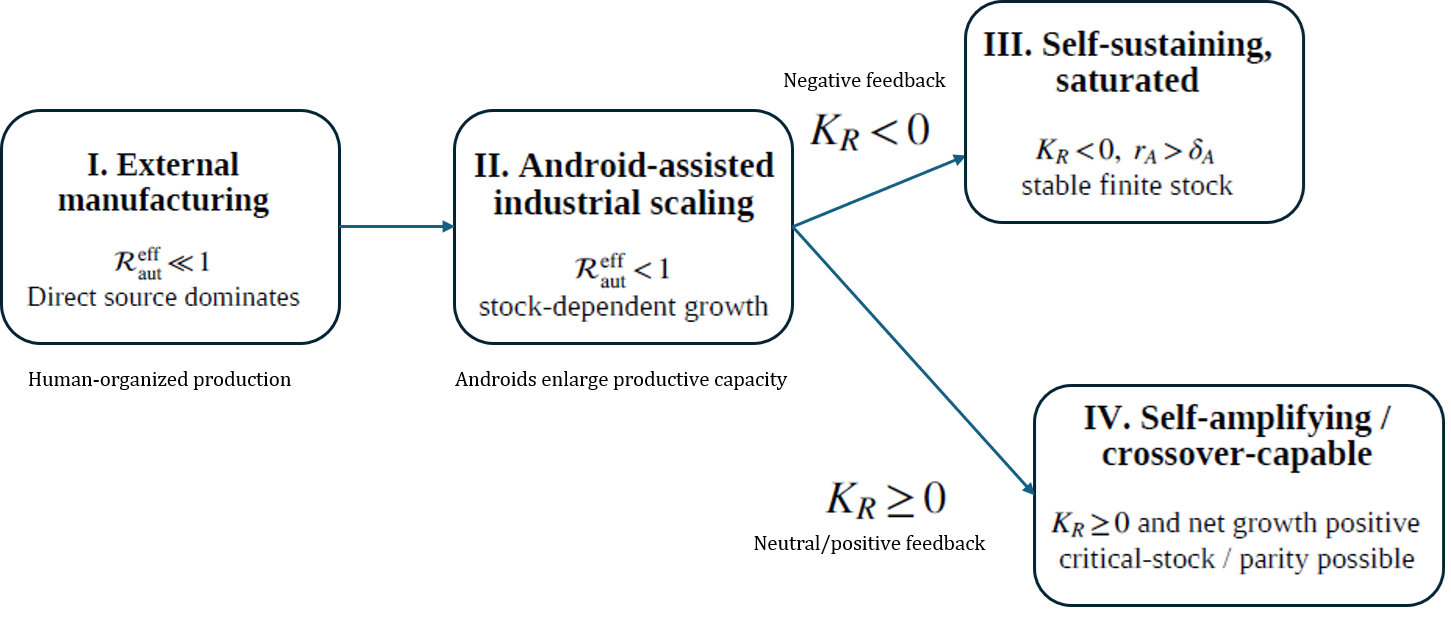}
\caption{Schematic dynamical classification of a technological population. The system progresses from externally manufactured androids to an android-assisted industrial regime as the installed stock becomes economically significant. If android-mediated production becomes dominant, negative feedback leads to a self-sustaining but saturated population, whereas neutral or positive feedback can produce a self-amplifying, crossover-capable regime. Production autonomy and numerical parity are distinct transitions.}
\label{fig:regimes}
\end{figure}

The analytical results suggest four qualitatively distinct regimes, which are schematically summarized in Fig.~\ref{fig:regimes}. The boundaries between them are not all demographic boundaries: some concern the production system itself.

\paragraph{Regime I: externally manufactured population.}
The Direct source and conventional industrial channel dominate. Quantitatively,
\begin{equation}
\mathcal{R}_{\rm aut}^{\rm eff}\ll1.
\end{equation}
The technological population exists only because the external human-organized production system continues to manufacture it.

\paragraph{Regime II: android-assisted industrial scaling.}
The stock of androids has become large enough to affect productive capacity, logistics, maintenance, or factory operation, but the gross android-mediated channel remains smaller than the external baseline:
\begin{equation}
0<\mathcal{R}_{\rm aut}^{\rm eff}<1.
\end{equation}
This regime can be economically important even though it is not production-autonomous.

\paragraph{Regime III: self-sustaining but saturated technological population.}
For $K_R<0$ and $r_A>\delta_A$, the isolated autonomous subsystem possesses the stable finite fixed point Eq.~(\ref{eq:stable_saturation}). Android-mediated production can exceed losses and even dominate the production system, but negative feedback limits the long-time stock. Whether numerical parity with humans occurs depends on whether that stable stock, together with the baseline channel, is sufficiently large.

\paragraph{Regime IV: self-amplifying, crossover-capable technological population.}
For $K_R=0$ with positive net growth, or for $K_R>0$ above the critical-stock threshold, the autonomous channel does not possess a stable finite fixed point in the minimal subsystem. The stock can therefore accelerate strongly until some omitted physical constraint becomes important. Numerical parity is possible in this regime but is not automatic: the integrated condition Eq.~(\ref{eq:crossover}) must still be satisfied.

\section{Discussion}
\label{sec:discussion}

The central result is a hierarchy of conditions rather than a single date. A billion-unit humanoid stock by 2050, if realized, is still approximately one order of magnitude smaller than the human population. Under the damped baseline calibrated to the present scenario anchors, that gap does not close by 2100. Numerical crossover requires a qualitatively stronger stock-dependent production channel, and the phase diagram quantifies how its required magnitude depends on nonlinear feedback.

The distinction between Direct and Replicative growth is therefore more than semantic. Present humanoid production is Direct in the sense of Ref.~\cite{TunstallZaccone2026}: an existing robot does not need to reproduce in order for a new robot to be manufactured. The calibrated $(q_M+r_Ma)e^{K_Ma^\gamma}$ term describes the effective external adoption channel. By contrast, $r_Aa e^{K_Ra}$ is derived from the canonical TZ productivity-feedback logic and is reserved for a future regime in which the installed android stock itself contributes measurably to the capacity that creates additional androids. The two nonlinear factors consequently refer to different observables and different physical mechanisms; keeping $K_M$ and $K_R$ distinct prevents ordinary market growth from being mislabeled as autonomous reproduction and avoids double counting the same feedback.

The production-autonomy threshold and the demographic crossover are independent conditions. Equations~(\ref{eq:autonomy_exact})--(\ref{eq:autonomy_approx}) ask when robot-mediated production exceeds conventional production. Equation~(\ref{eq:crossover}) asks whether the cumulative android per-capita growth advantage is large enough to erase the human/android numerical gap. Production autonomy can therefore occur while androids remain a small numerical minority, or numerical parity could in principle occur through massive conventional manufacturing without a fully autonomous production chain.

A second distinction concerns numerical and functional crossover. Economic importance may occur at much smaller android populations if the average useful annual output per robot exceeds that per human worker. If $L_H$ denotes the human labour force and $\varepsilon_H$, $\varepsilon_A$ denote effective annual useful outputs, a functional crossover would satisfy
\begin{equation}
\varepsilon_A A>\varepsilon_H L_H.
\label{eq:functional_crossover}
\end{equation}
Equation~(\ref{eq:functional_crossover}) is intentionally not calibrated here because there is not yet a sufficiently robust empirical basis for the lifetime productivity of general-purpose humanoids. It nevertheless shows why economic dominance, numerical parity, and production autonomy are three different thresholds.

The 2050 stress test is useful precisely because it is extreme. It demonstrates that a headline date can be translated into a required dynamical rate. The required $73.5\%$ compound annual increase from the 2025 stock to numerical parity in 2050 is far above the growth rate needed for parity by 2100. Within the nonlinear model, switching on a strong positive-feedback channel in 2035 lowers the required net coefficient relative to neutral feedback, but the coefficient remains very large. Thus a 2050 crossover is mathematically possible within the model only under assumptions that represent a profound change from the baseline production regime.

A further physical constraint is energy demand. A technological population approaching billions of units would require energy not only for locomotion and charging, but also for on-board computation, communication, maintenance, component fabrication, and the upstream AI and data-centre infrastructure on which many systems would depend. The scale of the latter is already non-negligible: the International Energy Agency projects global data-centre electricity demand to rise from roughly 485~TWh in 2025 to about 950~TWh in 2030 \cite{IEA2026EnergyAI}. Energy availability should therefore ultimately enter a more complete model as an explicit resource variable, or indirectly through stock-dependent changes in the effective feedback coefficients. This is another reason why the fastest positive-feedback trajectories should be interpreted as threshold diagnostics rather than unconstrained long-run forecasts.

An additional feature of the crossover problem is that the human reference population is itself non-monotonic. In the WPP2024 median trajectory, the global human population reaches a maximum of approximately $10.3\times10^9$ in the mid-2080s and subsequently begins a slow decline \cite{UNWPP2024}. Numerical crossover can therefore be approached through two simultaneous effects: growth of the android population and, after ``Peak Human'', decline of the human denominator. The exact criterion in Eq.~(\ref{eq:crossover}) already incorporates both effects through the difference $g_A-g_H$; once $g_H<0$, a given positive android growth rate produces a larger relative-growth advantage than it would during the human-growth phase.

\subsection{A multi-species geometrical extension}
\label{subsec:multispecies}

As stated explicitly in Assumption A5, the present calculations deliberately omit cross-species network feedback from the calibrated baseline, while treating the human trajectory as exogenous and the android production channels as separately identifiable. A natural next step, suggested by the geometrical interpretation of Ref.~\cite{TunstallZaccone2026}, is to regard humans and androids as two component species of a common growth-supporting network. In such a description the relevant state variable need not be the abundance of either population separately, but the effective number of components contributing to the same productive or technological network.

Care is required in defining such a combined variable. The quantities $h$ and $a$ used above are normalized by different population scales and therefore should not simply be added. Introduce instead a common normalization $N_s$ and define
\begin{equation}
u=\frac{H}{N_s},
\qquad
v=\frac{A}{N_s}.
\label{eq:common_network_variables}
\end{equation}
For a generic network-controlled rate, the geometrical construction then admits several nested two-species closures:
\begin{align}
r_1(u,v)
&=
r_0(1+\epsilon_H)^{u^{\gamma_H}}
   (1+\epsilon_A)^{v^{\gamma_A}},
\label{eq:multispecies_1}
\\
r_2(u,v)
&=
r_0(1+\epsilon)^{u^{\gamma_H}+v^{\gamma_A}},
\label{eq:multispecies_2}
\\
r_3(u,v)
&=
r_0(1+\epsilon)^{u^\gamma+v^\gamma},
\label{eq:multispecies_3}
\\
r_4(u,v)
&=
r_0(1+\epsilon)^{(u+v)^\gamma}.
\label{eq:multispecies_4}
\end{align}
The first form describes two populations contributing through distinct interaction structures; the second assumes a common interaction strength but allows different geometrical scaling exponents; the third additionally imposes the same exponent; and the fourth treats humans and androids as components of a single network before the geometrical scaling is applied.

The last case is particularly interesting because, for $\gamma\neq1$,
\begin{equation}
(u+v)^\gamma
\neq
u^\gamma+v^\gamma.
\end{equation}
It is therefore intrinsically non-separable and contains genuine mixed human--android contributions. For example, at $\gamma=2$,
\begin{equation}
(u+v)^2=u^2+2uv+v^2,
\end{equation}
so that the cross term $2uv$ represents interactions that exist only because both component species are present. In exponential notation, Eq.~(\ref{eq:multispecies_4}) is equivalent to
\begin{equation}
r_4(u,v)
=
r_0\exp\!\left[K(u+v)^\gamma\right],
\qquad
K=\ln(1+\epsilon),
\label{eq:multispecies_TZ}
\end{equation}
which provides a direct multi-species geometrical extension of the TZ rate-feedback structure.

A further generalization can represent substitution of android activity for human activity within the same network. Define an effective network coordinate
\begin{equation}
n_{\rm eff}(t)
=
g_H(t)u+\omega_A v,
\label{eq:effective_network_coordinate}
\end{equation}
where $g_H(t)$ measures the fraction or effectiveness of the human population participating in the relevant productive network and $\omega_A$ measures the relative contribution of one android component. The corresponding rate
\begin{equation}
r_{\rm net}(t)
=
r_0\exp\!\left[K n_{\rm eff}(t)^\gamma\right]
\label{eq:effective_network_rate}
\end{equation}
allows human participation to decrease even while the human headcount remains large, as would occur if androids progressively replace functions previously performed by humans.

The simple cross-coupled closure anticipated by the present model is recovered as a linear-state separable limit. For example,
\begin{equation}
\dot h
=
r_H h\exp(K_Hh+J_{HA}a),
\label{eq:human_crosscoupled}
\end{equation}
where $J_{HA}>0$ could represent beneficial demographic feedback from productivity, care, or health technologies, whereas $J_{HA}<0$ could represent adverse socioeconomic effects. An analogous term $J_{AH}h$ can be introduced in the android equation. The present data are insufficient to distinguish among Eqs.~(\ref{eq:multispecies_1})--(\ref{eq:multispecies_4}) or to estimate the activity weights in Eq.~(\ref{eq:effective_network_coordinate}); they should therefore be regarded as a falsifiable class of future extensions rather than additional fitted parameters of the current model.

Other extensions are also natural. Materials, chips, capital, and other bottleneck resources can be represented explicitly rather than being absorbed into $K_M$ and $K_R$. The global mean-field model can be generalized to regions or network nodes using the flux-coupled extensions proposed for the TZ framework \cite{ZacconeEtAl2026Review}, and stochastic forcing could represent technological breakthroughs, regulatory shocks, or supply disruptions. Such extensions should be introduced only when data justify their additional parameters. The minimal model developed here is intentionally low-dimensional so that its thresholds remain interpretable.

\section{Limitations}
\label{sec:limitations}

The main limitation concerns android data. The calibration is a scenario calibration, not an empirical population fit. The 2025 stock is an approximate estimate, the 2050 stock is a market-research projection, and the intermediate 2030--2050 proxy points inherit the shape of a U.S. scenario. The improved generalized-TZ fit is therefore evidence of functional adequacy for that assumed timing curve, not evidence of predictive accuracy for future global deployments.

Second, the parameters are assumed constant over each scenario window. The fitted $(K_M,\gamma)$ pair should therefore be interpreted as an average effective description of the external adoption channel over 2025--2050. Changes in unit cost, production technology, regulation, demand, resource availability, or AI capability would make $K_M$, $\gamma$, $r_M$, or $q_M$ time dependent. The hypothetical autonomous parameter $K_R$ is conceptually different and is not estimated from these data at all.

Third, as stated in Assumption A5, the baseline model omits explicit human--android network coupling: WPP2024 supplies the human trajectory independently of android abundance, and the additional autonomous-production channel is driven by android stock rather than an unidentified combined-network variable. If humanoids materially affect fertility, mortality, migration, wealth distribution, care, conflict, or the human contribution to technological production, the coupled trajectories would change. The multi-species construction in Sec.~\ref{subsec:multispecies} provides a compact route toward such coupling, although a detailed demographic implementation would require additional data.

Fourth, retirement is represented by a constant hazard. Real machines have warranty periods, maintenance cycles, upgrades, and age-dependent failure, so an age-structured stock-flow model would ultimately be preferable. Likewise, an android-mediated production coefficient does not distinguish mining, energy generation, component fabrication, chip production, assembly, software, and logistics. Full industrial autonomy would require closure of several linked production chains, not merely a large value of one coarse-grained coefficient.

Fifth, the positive-feedback model can produce extremely rapid growth when $K_R>0$. Such trajectories indicate that the minimal model has entered a regime in which omitted constraints must eventually matter. They should not be extrapolated arbitrarily far. Resource depletion, market saturation, physical manufacturing limits, and policy responses would regularize the dynamics.

Finally, numerical crossover is not equivalent to social, political, economic, or technological dominance. The mathematically defined event is only $A=H$. The terms ``stress test'' and ``crossover'' are used to maintain this distinction.

\section{Conclusions}

The sharpest model prediction is the dimensionless transition
\begin{equation}
\mathcal{P}(a)=\frac{r_A}{\delta_A}e^{K_Ra}=1.
\end{equation}
This result follows from applying the canonical TZ productivity-feedback construction directly to robot-mediated production: the gross productivity per active android is $r_R(a)=r_Ae^{K_Ra}$. Below $\mathcal{P}=1$ retirement dominates the isolated autonomous channel; above it the channel is locally self-amplifying. The fitted external-adoption feedback $(K_M,\gamma)$ is a separate empirical coarse-graining and is not identified with $K_R$. The generalized baseline is strongly preferred to the canonical $\gamma=1$ baseline and to endpoint-constrained exponential, logistic and Gompertz alternatives within the adopted forecast-derived scenario. A global uncertainty analysis confirms that precise crossover dates remain strongly
scenario-dependent; the more robust prediction is the production--feedback threshold
itself. Detailed uncertainty and sensitivity analysis results are provided in the Appendix.

We have formulated a minimal nonlinear model for the coexistence of biological and technological populations. The framework combines the TZ state-dependent rate with the Direct/Replicative distinction of geometrical growth models. Calibrated to the UN human trajectory and a published humanoid-adoption scenario, the baseline android model reaches about $3.96\times10^9$ units by 2100 and does not overtake the human population. Numerical crossover requires an additional population-proportional android-mediated production channel.

For neutral autonomous feedback the critical net coefficient for parity by 2100 is approximately 0.030 yr$^{-1}$ in the present calibration. Positive nonlinear feedback lowers the threshold, while negative feedback raises it. A deliberately extreme early-autonomy stress test shows that a 2050 numerical crossover would require much stronger dynamics: if the additional channel switches on in 2035, the threshold is $\mu_c\simeq0.269$ yr$^{-1}$ for $K_R=0$ and $\mu_c\simeq0.205$ yr$^{-1}$ for $K_R=0.10$. These are conditional stress-test boundaries, not technological forecasts.

The model suggests a practical empirical program. As humanoid deployment expands, one should measure not only the installed stock but also gross production, retirement, productive hours, the fraction of robot manufacturing performed by robots, and resource intensity. Those observables would allow the Direct and Replicative channels to be estimated separately and would determine whether the human--android system is moving toward saturation, coexistence, or numerical crossover.

\section*{Data availability}

The human population data are from the United Nations World Population Prospects 2024. The humanoid numbers used here are published scenario estimates and forecast anchors cited in the text. The numerical calibration values, generalized-TZ parameters, benchmark table, and threshold tables used to generate the figures are supplied with the manuscript source. The full uncertainty and sensitivity analysis is provided in the Appendix.

\section*{Declaration of competing interest}

The authors declare that they have no known competing financial interests or personal relationships that could have appeared to influence the work reported in this paper.

\section*{Acknowledgements}

The authors thank Joseph Chamie for his demographic essays on androids in human populations. ChatGPT Plus was used to polish the text and to cross-check calculations and results.

\clearpage
\section*{Appendix: Supplementary Material}
\addcontentsline{toc}{section}{Appendix: Supplementary Material}

This Appendix contains the uncertainty and sensitivity analysis supporting the main paper,
together with the full pedagogical derivation connecting the Trachenko--Zaccone growth
equation to classical population-growth laws.

\setcounter{section}{0}
\setcounter{subsection}{0}
\setcounter{equation}{0}
\setcounter{figure}{0}
\setcounter{table}{0}

\renewcommand{\thesection}{S\arabic{section}}
\renewcommand{\thesubsection}{S\arabic{section}.\arabic{subsection}}
\renewcommand{\theequation}{S\arabic{equation}}
\renewcommand{\thefigure}{S\arabic{figure}}
\renewcommand{\thetable}{S\arabic{table}}

\makeatletter
\renewcommand{\theHsection}{S\arabic{section}}
\renewcommand{\theHsubsection}{S\arabic{section}.\arabic{subsection}}
\renewcommand{\theHequation}{S\arabic{equation}}
\renewcommand{\theHfigure}{S\arabic{figure}}
\renewcommand{\theHtable}{S\arabic{table}}
\makeatother

\section{Construction of the uncertainty ensemble}
\label{sec:s1}

The android component is forecast-dominated, so a single parameter set would give a misleading impression of precision. We therefore vary five poorly known inputs simultaneously. The adopted ranges are deliberately broad epistemic envelopes rather than statistical confidence intervals:
\begin{eqnarray}
10^3 &\leq& A(2025) \leq 10^5,\nonumber\\
3\times10^8 &\leq& A(2050) \leq 2\times10^9,\nonumber\\
2030 &\leq& t_a \leq 2050,\nonumber\\
5~{\rm yr} &\leq& L \leq 20~{\rm yr},\nonumber\\
-0.10 &\leq& K_R \leq 0.10.
\label{eq:uncertainty_ranges}
\end{eqnarray}
Here $L=1/\delta_A$ is the mean effective service lifetime and $t_a$ is the onset year of android-mediated production. Morgan Stanley's detailed market analysis uses an eight-year replacement cycle as one working assumption \cite{MorganStanleyHumanoid100}; the wider 5--20 yr range is used here to test sensitivity rather than to assert a probability distribution.

We draw $N=4096$ points using Latin-hypercube sampling \cite{McKay1979}. The 2025 and 2050 population endpoints are sampled log-uniformly because their uncertainty is naturally multiplicative. The onset year, service lifetime, and feedback parameter are sampled uniformly over the intervals in Eq.~(\ref{eq:uncertainty_ranges}). The resulting ensemble is a design for uncertainty propagation, not a posterior distribution.

\section{Propagation of endpoint uncertainty to the autonomy-onset stock}
\label{sec:s2}

Changing the 2025 and 2050 endpoints should not arbitrarily distort the assumed timing of humanoid adoption. We therefore preserve the reference logarithmic progress along the forecast-derived adoption trajectory. Let $A_{\rm ref}(t)$ denote the reference timing curve and define
\begin{equation}
u(t)=
\frac{\ln A_{\rm ref}(t)-\ln A_{\rm ref}(2025)}
{\ln A_{\rm ref}(2050)-\ln A_{\rm ref}(2025)}.
\label{eq:progress_function}
\end{equation}
By construction, $u(2025)=0$ and $u(2050)=1$. For a sampled endpoint pair $[A_0,A_{2050}]$, the stock at the autonomy-onset year is obtained by linear interpolation on the logarithmic population scale,
\begin{equation}
\ln A(t_a)
=
[1-u(t_a)]\ln A_0
+
u(t_a)\ln A_{2050}.
\label{eq:onset_stock}
\end{equation}
Exponentiating gives
\begin{equation}
A(t_a)
=
A_0^{1-u(t_a)}A_{2050}^{u(t_a)}.
\label{eq:onset_stock_explicit}
\end{equation}
Thus endpoint uncertainty changes the scale of the trajectory while preserving its reference temporal progress.

\section{Critical coefficient for parity by 2100}
\label{sec:s3}

For the robustness calculation we deliberately remove the conventional externally organized manufacturing term after $t_a$ and retain only the autonomous life-cycle subsystem. This is the canonical TZ-derived robot-mediated channel of the main text, with per-unit gross productivity $r_R(a)=r_Ae^{K_Ra}$,
\begin{equation}
\dot a
=
a\left[r_Ae^{K_Ra}-\frac{1}{L}\right].
\label{eq:autonomous_uncertainty}
\end{equation}
This choice is conservative: any continued external manufacturing would reduce the gross autonomous coefficient required for parity.

Throughout this calculation we retain the main-text normalizations $a=A/A_s$ with $A_s=10^9$ and $h=H/H_s$ with $H_s=10^{10}$. Thus $a(t_a)=A(t_a)/A_s$. The critical value $r_{A,c}$ is defined as the value that gives numerical parity at the end of the horizon,
\begin{equation}
a(2100)=\frac{H_s}{A_s}h(2100)=10h(2100),
\label{eq:parity_condition}
\end{equation}
which is equivalent to the physical condition $A(2100)=H(2100)$. Dividing Eq.~(\ref{eq:autonomous_uncertainty}) by $a$ gives
\begin{equation}
\frac{1}{a}\frac{da}{dt}
=
r_Ae^{K_Ra}-\frac{1}{L}.
\end{equation}
With the substitution
\begin{equation}
z=\ln a,
\qquad
\frac{dz}{dt}=\frac{1}{a}\frac{da}{dt},
\end{equation}
we obtain
\begin{equation}
\frac{dz}{dt}
=
r_A\exp(K_Re^z)-\frac{1}{L}.
\end{equation}
Separating variables between $t_a$ and 2100 yields
\begin{equation}
2100-t_a
=
\int_{\ln a(t_a)}^{\ln [10h(2100)]}
\frac{dz}
{r_{A,c}\exp(K_Re^z)-1/L}.
\label{eq:critical_integral}
\end{equation}
For every Latin-hypercube sample, Eq.~(\ref{eq:critical_integral}) is solved numerically for $r_{A,c}$. The denominator is required to remain positive along the relevant trajectory; otherwise the sampled autonomous channel cannot reach parity monotonically under the assumed minimal subsystem.

It is also useful to report the dimensionless gross production-to-loss ratio
\begin{equation}
\Pi_c
=
\frac{r_{A,c}}{\delta_A}
=
r_{A,c}L.
\label{eq:Pi_c}
\end{equation}
This quantity connects directly to the production--feedback number introduced in the main text.

\section{Distribution of crossover thresholds}
\label{sec:s4}

Across the 4096-sample uncertainty ensemble, the critical gross coefficient has median
\begin{equation}
{\rm median}(r_{A,c})=0.178~{\rm yr}^{-1},
\end{equation}
with interquartile range
\begin{equation}
0.142<r_{A,c}<0.218~{\rm yr}^{-1}
\end{equation}
and 95\% design interval
\begin{equation}
0.095<r_{A,c}<0.340~{\rm yr}^{-1}.
\end{equation}
The corresponding dimensionless ratio $\Pi_c=r_{A,c}/\delta_A$ has median $2.08$ and 95\% design interval $1.21<\Pi_c<3.44$.

\begin{figure}[t]
\centering
\includegraphics[width=0.84\linewidth]{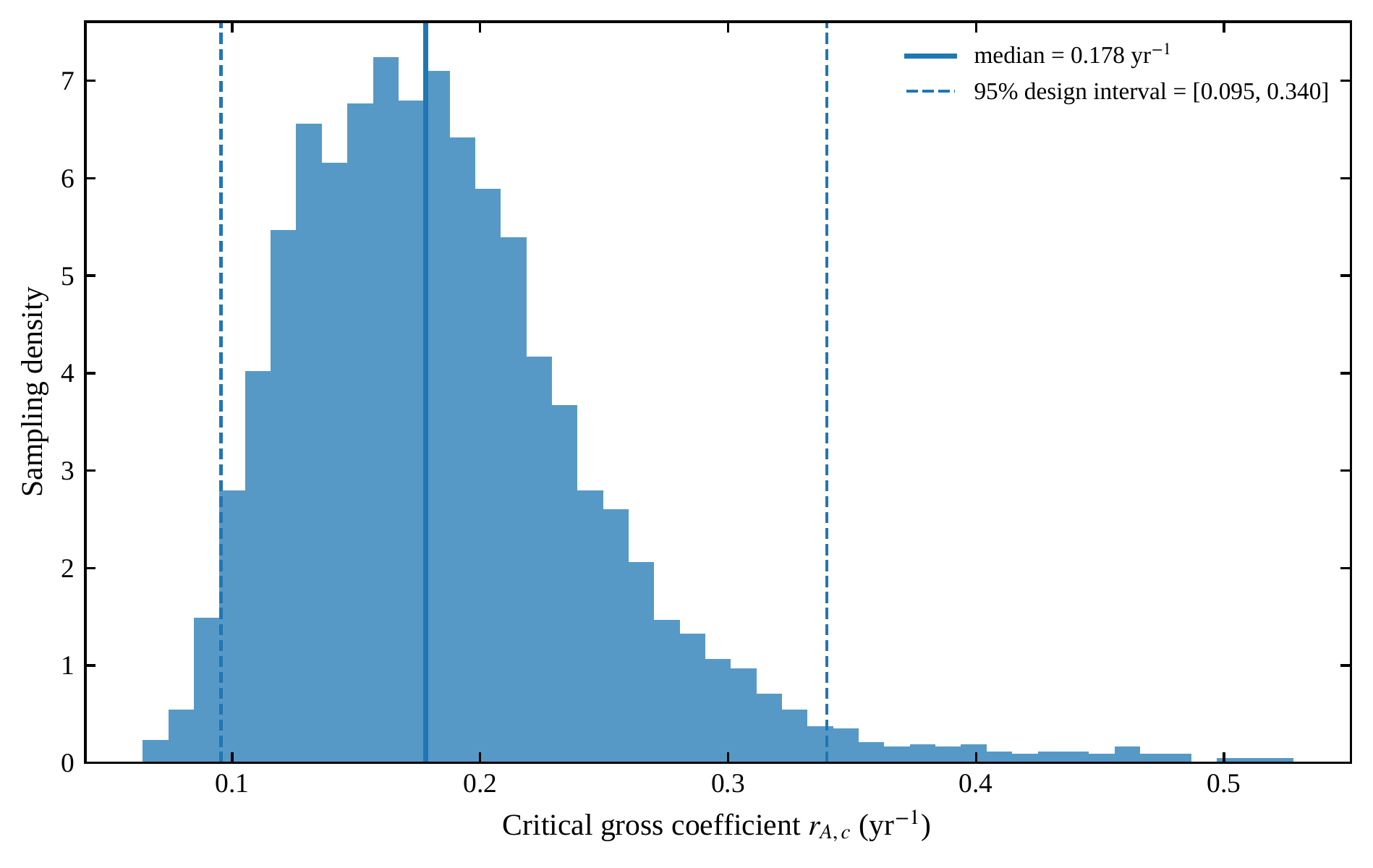}
\caption{Distribution of the critical gross android-mediated production coefficient required for parity by 2100 across 4096 Latin-hypercube samples. Dashed lines mark the 2.5th and 97.5th percentiles. This is an uncertainty-design distribution, not a posterior probability distribution.}
\label{fig:uncertainty_distribution}
\end{figure}

The broad interval demonstrates why a single crossover date should not be interpreted as a robust forecast. What remains more stable is the qualitative requirement that gross android-mediated production must exceed effective retirement by a sufficient margin, with the required margin modified by the nonlinear feedback $K_R$.

\section{Global rank sensitivity analysis}
\label{sec:s5}

To identify which uncertain assumptions control the threshold most strongly, we compute Spearman rank correlations between each sampled input and $r_{A,c}$. The results are listed in Table~\ref{tab:sensitivity}.

\begin{table}[t]
\centering
\caption{Global rank sensitivity of the critical gross android-mediated production coefficient $r_{A,c}$.}
\label{tab:sensitivity}
\begin{tabular}{lc}
\toprule
Input & Spearman $\rho$ with $r_{A,c}$\\
\midrule
Mean service lifetime $L$ & -0.677\\
Autonomy-onset year $t_a$ & -0.449\\
Feedback $K_R$ & -0.443\\
$\log_{10}A(2050)$ & -0.118\\
$\log_{10}A(2025)$ & -0.092\\
\bottomrule
\end{tabular}
\end{table}

\begin{figure}[t]
\centering
\includegraphics[width=0.82\linewidth]{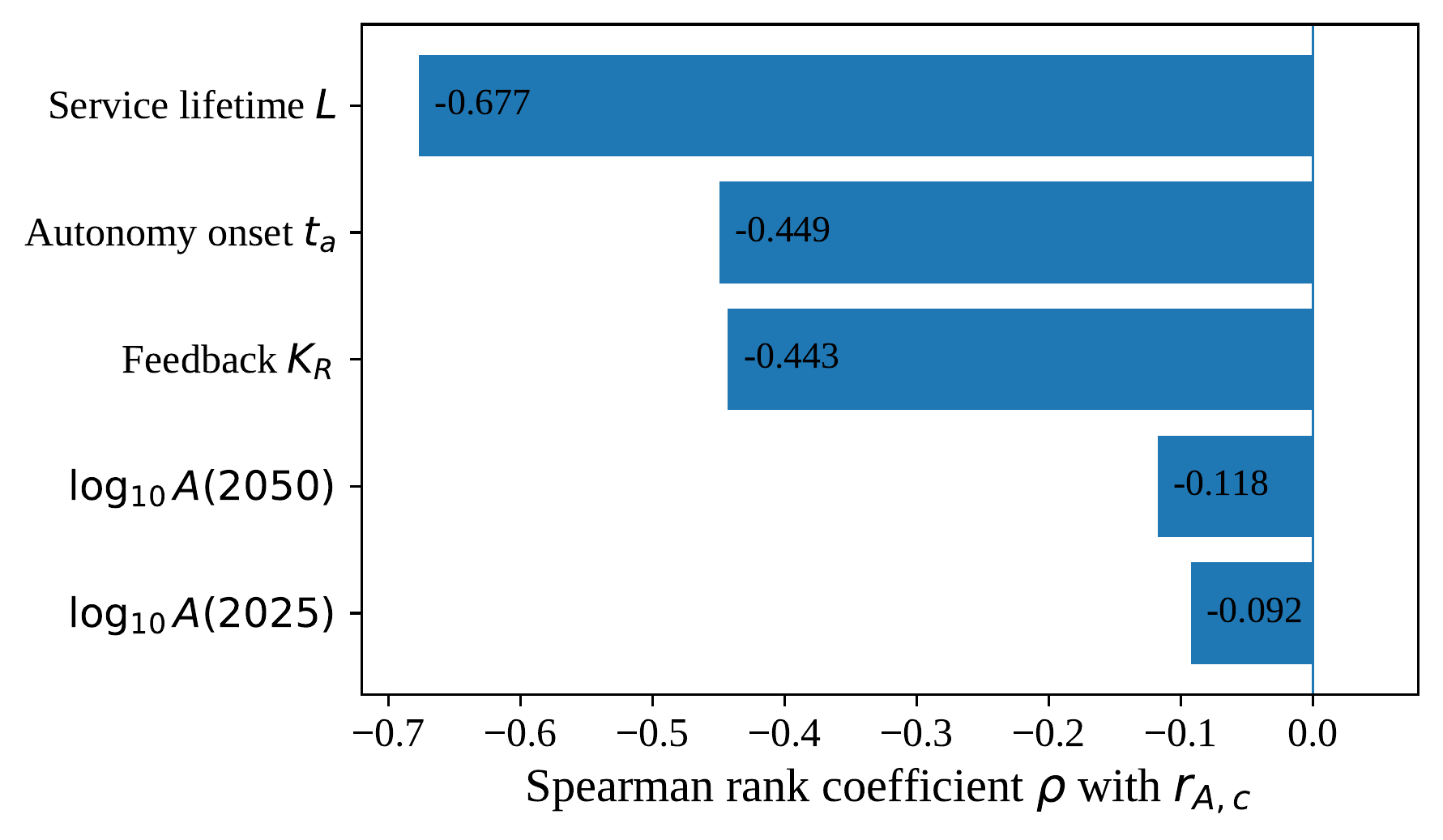}
\caption{Spearman rank sensitivity of the critical gross android-mediated coefficient $r_{A,c}$ to the five uncertain inputs. The magnitude of $\rho$ measures the strength of the monotonic influence over the adopted uncertainty design; the sign gives the direction of that influence.}
\label{fig:rank_sensitivity}
\end{figure}

The strongest sensitivity is to service lifetime. A longer-lived android stock requires a smaller gross production coefficient because fewer units must be replaced. The next strongest dependencies are the autonomy-onset year and $K_R$. The negative onset-year coefficient may appear counterintuitive because later onset leaves less time before 2100. In the adopted uncertainty design, however, a later onset also occurs much farther up the assumed adoption trajectory, so the larger starting stock more than compensates for the shorter remaining time. Positive $K_R$ similarly lowers the gross base coefficient because the feedback factor increases as the stock grows.

The two endpoint uncertainties are appreciably weaker over the adopted ranges. In particular, uncertainty in the exact 2025 humanoid count has little leverage on the threshold compared with lifetime and feedback. This result suggests that future empirical effort should prioritize measurements of installed-stock survival, retirement, gross robot-mediated production, and the stock dependence of production capacity rather than only the absolute early population count.

\section{Additional robustness interpretation}
\label{sec:s6}

The uncertainty analysis clarifies the distinction between three statements made in the main paper.

First, the numerical crossover year is scenario-dependent. It depends on the stock inherited at the onset of autonomy, the remaining time horizon, retirement, and nonlinear production feedback.

Second, the production-autonomy transition is a different event. It is governed by the dimensionless production--feedback number
\begin{equation}
\mathcal{P}(a)
=
\frac{r_A}{\delta_A}e^{K_Ra},
\end{equation}
and the local autonomous subsystem becomes self-amplifying at $\mathcal{P}=1$. This structural threshold is independent of the arbitrary choice of 2100 as a demographic comparison horizon.

Third, numerical parity requires the supercritical or sufficiently high-growth state to persist long enough to close the initial logarithmic human--android population gap. Hence uncertainty in the date of parity does not invalidate the existence of the underlying production--feedback threshold.

The numerical files supplied with the manuscript include the complete 4096-point Latin-hypercube sample, the critical coefficients, and the rank-sensitivity outputs so that the analysis can be reproduced or updated as new data become available.

\section*{Supplementary data files}
The following machine-readable files accompany the source package:
\begin{itemize}
\item \texttt{uncertainty\_lhs\_4096.csv},
\item \texttt{sensitivity\_spearman.csv},
\item \texttt{stress\_test\_2050\_thresholds.csv},
\item \texttt{model\_benchmark.csv}.
\end{itemize}


\section{Pedagogical derivation: from the Trachenko--Zaccone equation to classical population-growth laws}
\label{sec:s7_pedagogical_TZ}
The Trachenko--Zaccone equation was originally derived in the context of time relaxation of physical properties of amorphous materials, such as glasses and supercooled liquids, based on local-event dynamics in disordered systems. At the microscopic level, elastically interacting rearrangement events of atomic or molecular nearest-neighbour cages (local molecular environments) either trigger (favour) new events or make them unfavourable. In the former situation, a feed-forward mechanism leads to avalanche-like dynamics characteristic of glass rejuvenation \cite{Ruta2012MetallicGlass}, whereas in the latter case a feedback mechanism leads to sluggish ageing-type dynamics \cite{TrachenkoZaccone2021}. The resulting population balance of local events leads to a simple nonlinear differential equation, the TZ equation, whose solutions closely approximate stretched-exponential (Kohlrausch), compressed-exponential, simple exponential (as in high-temperature liquids) and even logarithmic relaxation of physical properties (dielectric response, viscoelastic moduli, enthalpy or volume, etc.) \cite{GinzburgGendelmanZaccone2024,ZacconeGinzburgGendelmanMauroMauro2026}.
Flipping the sign of the right-hand side of the TZ equation turns a successful model of glassy relaxation in condensed matter into a mathematical model of growth, as the solutions are now growing functions of time $t$. In a recent contribution \cite{ZacconeTrachenko2026}, this population-growth version of the TZ equation was shown to describe the past 12,000 years of global population growth, from the Neolithic to the present. 

Before applying the Trachenko--Zaccone (TZ) framework for population growth to the
human--android problem, we derive its population form and show how several classical laws of
population dynamics emerge as limiting cases.

\subsection{From a population balance to a state-dependent growth rate}

Consider a population containing $N(t)$ elements. Its rate of change
may always be written in the population-balance form

\begin{equation}
\frac{dN}{dt}
=
r(N)N,
\label{eq:general_population_balance}
\end{equation}

where $r(N)$ is the net productivity per population element. In a
biological population, $r$ may be interpreted as the difference
between per-capita birth and death rates. In a technological
population, it can instead represent the net production rate per
existing unit. In human demography, this birth--death decomposition can itself be analysed through fertility and survival components, a perspective that has been fruitfully combined with logistic descriptions of population change \cite{MarchettiMeyerAusubel1996}.

The classical Malthus model assumes that $r$ is constant. The central
idea of the TZ framework is to relax this assumption and allow the
instantaneous rate itself to depend on the current state of the
system. A biological analogue of such state-dependent collective regulation is bacterial quorum sensing, in which chemical signalling enables cells to detect population density and switch group behaviours once density-dependent signals accumulate \cite{WatersBassler2005}. The analogy is only structural: the microscopic mechanism is entirely different, but in both cases the current population state modifies the effective activity of its constituents.

In the original derivation of the TZ equation for relaxation,
interactions between successive local events modify an effective
activation barrier. Since the elementary transition probability
depends exponentially on that barrier, a barrier that varies linearly
with the state variable generates an exponential state dependence of
the rate. The same mathematical principle can be transferred to a
growing population by assuming that the logarithm of the effective
productivity changes linearly with population.

We introduce a dimensionless population

\begin{equation}
y(t)
=
\frac{N(t)}{N_s},
\label{eq:normalized_population}
\end{equation}

where $N_s$ is a convenient population scale. We then write the
state-dependent productivity as

\begin{equation}
r(y)
=
\frac{1}{\tau}
\exp(Ky),
\label{eq:TZ_productivity}
\end{equation}

where $\tau$ sets the characteristic time scale and $K$ is a
dimensionless feedback parameter. Substitution into
Eq.~(\ref{eq:general_population_balance}) gives

\begin{equation}
\frac{dy}{dt}
=
\frac{y}{\tau}
\exp(Ky).
\label{eq:TZ_population}
\end{equation}

Equation~(\ref{eq:TZ_population}) is the growth counterpart of the
usual TZ relaxation equation. The difference is simply the overall
sign: relaxation decreases the relevant state variable, whereas a
population-growth process increases it
\cite{ZacconeTrachenko2026,ZacconeEtAl2026Review}.

The physical meaning of $K$ is immediately apparent from the
instantaneous per-capita growth rate,

\begin{equation}
\frac{1}{y}\frac{dy}{dt}
=
\frac{1}{\tau}\exp(Ky).
\label{eq:percapita_TZ}
\end{equation}

For $K<0$, productivity decreases as the population grows and the
feedback is inhibitory. For $K=0$, productivity is independent of
population. For $K>0$, productivity increases with population and the
feedback is self-amplifying.

\subsection{A useful low-order expansion}

The relation between Eq.~(\ref{eq:TZ_population}) and classical
population models becomes particularly transparent by expanding the
exponential,

\begin{equation}
\exp(Ky)
=
1
+
Ky
+
\frac{K^2y^2}{2}
+
\frac{K^3y^3}{6}
+
\cdots .
\label{eq:TZ_expansion}
\end{equation}

Substitution into Eq.~(\ref{eq:TZ_population}) gives

\begin{equation}
\frac{dy}{dt}
=
\frac{y}{\tau}
+
\frac{K}{\tau}y^2
+
\frac{K^2}{2\tau}y^3
+
\cdots .
\label{eq:TZ_hierarchy}
\end{equation}

Equation~(\ref{eq:TZ_hierarchy}) provides a useful way of understanding
the TZ model. The familiar classical laws correspond to retaining
different low-order terms of a more general nonlinear rate-feedback
equation.

\subsection{Malthusian growth: the zero-feedback limit}

The simplest case is

\begin{equation}
K=0.
\end{equation}

Equation~(\ref{eq:TZ_population}) then becomes exactly

\begin{equation}
\frac{dy}{dt}
=
\frac{y}{\tau},
\label{eq:Malthus_from_TZ}
\end{equation}

which is the Malthus growth equation \cite{Malthus1798}. Integrating,

\begin{equation}
\frac{dy}{y}
=
\frac{dt}{\tau},
\end{equation}

and applying the initial condition $y(t_0)=y_0$ gives

\begin{equation}
y(t)
=
y_0
\exp\left(
\frac{t-t_0}{\tau}
\right).
\label{eq:Malthus_solution}
\end{equation}

Thus, ordinary exponential growth is not a separate model within this
framework: it is the exact zero-feedback limit of the TZ equation.

\subsection{Verhulst logistic growth: weak negative feedback}

Consider now

\begin{equation}
K<0
\end{equation}

and define

\begin{equation}
s=-K>0.
\end{equation}

For sufficiently weak nonlinearity,

\begin{equation}
|Ky|\ll 1,
\label{eq:weak_feedback_condition}
\end{equation}

the exponential may be truncated after its linear term,

\begin{equation}
\exp(Ky)
\simeq
1+Ky
=
1-sy.
\end{equation}

Equation~(\ref{eq:TZ_population}) therefore becomes

\begin{equation}
\frac{dy}{dt}
\simeq
\frac{y}{\tau}
\left(
1-sy
\right).
\label{eq:logistic_from_TZ_1}
\end{equation}

Introducing

\begin{equation}
y_c
=
\frac{1}{s}
=
-\frac{1}{K},
\label{eq:effective_carrying_capacity}
\end{equation}

we obtain

\begin{equation}
\frac{dy}{dt}
\simeq
\frac{y}{\tau}
\left(
1-\frac{y}{y_c}
\right),
\label{eq:logistic_from_TZ}
\end{equation}

which is precisely the mathematical form of the Verhulst logistic
equation \cite{Verhulst1838}.

Its solution is

\begin{equation}
y(t)
=
\frac{y_c}
{
1+
\left(
\frac{y_c}{y_0}-1
\right)
\exp\left[
-\frac{t-t_0}{\tau}
\right]
}.
\label{eq:logistic_solution}
\end{equation}

Negative TZ feedback causes the effective productivity to decrease as the
population grows. At first order this produces the familiar carrying-capacity
term of logistic growth.

This controlled local limit should not be confused with an exact global equivalence. Equation~(\ref{eq:logistic_from_TZ})
follows from the approximation $|Ky|\ll1$. The finite carrying
capacity $y_c=-1/K$ is therefore a property of the first-order
truncation and is not a non-zero fixed point of the full exponential
TZ equation.

\subsection{Positive feedback and the von Foerster hyperbolic law}

The opposite case,

\begin{equation}
K>0,
\end{equation}

corresponds to positive feedback: the per-capita productivity
increases as the population increases.

Again taking the controlled limit $|Ky|\ll1$,

\begin{equation}
\exp(Ky)
\simeq
1+Ky,
\end{equation}

gives

\begin{equation}
\frac{dy}{dt}
\simeq
\frac{y}{\tau}
+
\frac{K}{\tau}y^2.
\label{eq:Riccati_TZ}
\end{equation}

This is a Riccati-type growth equation. To solve it, we write

\begin{equation}
\frac{dy}
{y(1+Ky)}
=
\frac{dt}{\tau}.
\label{eq:Riccati_separation}
\end{equation}

Using

\begin{equation}
\frac{1}{y(1+Ky)}
=
\frac{1}{y}
-
\frac{K}{1+Ky},
\end{equation}

integration gives

\begin{equation}
\ln
\left(
\frac{y}{1+Ky}
\right)
=
\frac{t-t_0}{\tau}
+
\ln
\left(
\frac{y_0}{1+Ky_0}
\right).
\end{equation}

Solving for $y(t)$ yields

\begin{equation}
y(t)
=
\frac{
y_0
\exp[(t-t_0)/\tau]
}{
1-
Ky_0
\left\{
\exp[(t-t_0)/\tau]-1
\right\}
}.
\label{eq:Riccati_solution}
\end{equation}

For $K>0$, the denominator vanishes at the finite time

\begin{equation}
t_c
=
t_0
+
\tau
\ln
\left(
1+\frac{1}{Ky_0}
\right).
\label{eq:Riccati_critical_time}
\end{equation}

Close to this singular time, the leading behaviour of
Eq.~(\ref{eq:Riccati_solution}) is

\begin{equation}
y(t)
\simeq
\frac{\tau}
{K(t_c-t)}.
\label{eq:hyperbolic_limit}
\end{equation}

This is the hyperbolic form

\begin{equation}
y(t)
\simeq
\frac{A}{D-t},
\label{eq:vonFoerster_form}
\end{equation}

used by von Foerster, Mora and Amiot in their famous analysis of
historical world-population data \cite{vonFoerster1960}. Their
extrapolation led to the well-known ``Doomsday'' singularity in 2026.

The connection must be interpreted carefully. The von Foerster
hyperbola follows here from the first-order Riccati truncation of the
TZ equation. As the formal singularity is approached, $y$ becomes
large and the original condition $|Ky|\ll1$ eventually fails.
Consequently, Eq.~(\ref{eq:vonFoerster_form}) should be understood as
a controlled local correspondence rather than as the exact global
solution of the full TZ equation
\cite{ZacconeTrachenko2026}.

\subsection{A unified interpretation}
\label{subsec:unified_growth}

The different population-growth regimes discussed above can therefore be understood
within the same nonlinear rate-feedback equation,
\begin{equation}
\frac{dy}{dt}
=
\frac{y}{\tau}e^{Ky},
\label{eq:unified_growth}
\end{equation}
where the sign and magnitude of $K$ determine whether the instantaneous
per-capita productivity decreases, remains constant, or increases as the population
grows.

The instantaneous logarithmic growth rate is
\begin{equation}
g(y)
\equiv
\frac{1}{y}\frac{dy}{dt}
=
\frac{1}{\tau}e^{Ky}.
\label{eq:instantaneous_growth_rate}
\end{equation}
The three cases are therefore:
\begin{equation}
K<0
\qquad\Longrightarrow\qquad
\frac{dg}{dy}<0,
\end{equation}
corresponding to progressively inhibited growth,
\begin{equation}
K=0
\qquad\Longrightarrow\qquad
\frac{dg}{dy}=0,
\end{equation}
corresponding to ordinary exponential growth, and
\begin{equation}
K>0
\qquad\Longrightarrow\qquad
\frac{dg}{dy}>0,
\end{equation}
corresponding to progressively self-amplifying growth.

Over the relevant dynamical range, the solutions of the full nonlinear equation can
therefore be represented by generalized exponential growth laws of the form
\begin{equation}
y(t)
\sim
\exp\left[
\left(
\frac{t}{\tau_{\rm eff}}
\right)^{\beta}
\right],
\label{eq:generalized_exponential_growth}
\end{equation}
where the effective exponent $\beta$ distinguishes sub-exponential,
exponential, and super-exponential growth. In this sense, the full TZ equation
connects continuously the regimes
\begin{equation}
K<0
\qquad\Longrightarrow\qquad
0<\beta<1,
\qquad
\mbox{stretched-exponential growth},
\label{eq:stretched_growth}
\end{equation}

\begin{equation}
K=0
\qquad\Longrightarrow\qquad
\beta=1,
\qquad
\mbox{ordinary exponential growth},
\label{eq:exponential_growth}
\end{equation}

and
\begin{equation}
K>0
\qquad\Longrightarrow\qquad
\beta>1,
\qquad
\mbox{compressed-exponential growth}.
\label{eq:compressed_growth}
\end{equation}

The terminology ``stretched'' and ``compressed'' refers here to the growth curve
relative to ordinary exponential growth. For $K<0$, the per-capita productivity
decreases continuously as $y$ increases, so that the population grows progressively
more slowly than a simple exponential. For $K>0$, by contrast, the per-capita
productivity increases continuously with population, producing an accelerating,
super-exponential growth regime.

The classical population laws derived above appear as particularly transparent
limits of this more general structure. In the absence of feedback,
\begin{equation}
K=0
\qquad\Longrightarrow\qquad
\mbox{Malthusian exponential growth}.
\label{eq:malthus_unified}
\end{equation}

For weak negative feedback,
\begin{equation}
K<0,
\qquad
|Ky|\ll1,
\end{equation}
expansion of the exponential gives
\begin{equation}
e^{Ky}
\simeq
1+Ky,
\end{equation}
and therefore
\begin{equation}
\frac{dy}{dt}
\simeq
\frac{y}{\tau}
\left(
1-\frac{y}{y_c}
\right),
\qquad
y_c=-\frac{1}{K},
\end{equation}
which has the Verhulst logistic form. Thus,
\begin{equation}
K<0,\qquad |Ky|\ll1
\qquad\Longrightarrow\qquad
\mbox{Verhulst logistic-like growth}.
\label{eq:logistic_unified}
\end{equation}

Similarly, for weak positive feedback,
\begin{equation}
K>0,
\qquad
|Ky|\ll1,
\end{equation}
one obtains
\begin{equation}
\frac{dy}{dt}
\simeq
\frac{y}{\tau}
+
\frac{K}{\tau}y^2,
\label{eq:riccati_unified}
\end{equation}
which is a Riccati equation and leads to the hyperbolic finite-time behaviour
associated with the von Foerster population law. Hence,
\begin{equation}
K>0,\qquad |Ky|\ll1
\qquad\Longrightarrow\qquad
\mbox{Riccati/von Foerster-type growth}.
\label{eq:vonfoerster_unified}
\end{equation}

The hierarchy may therefore be summarized schematically as
\begin{equation}
\begin{array}{rcl}
K<0
&\Longrightarrow&
\mbox{stretched-exponential growth},\\[1mm]
K<0,\ |Ky|\ll1
&\Longrightarrow&
\mbox{Verhulst logistic-like growth},\\[1mm]
K=0
&\Longrightarrow&
\mbox{Malthusian exponential growth},\\[1mm]
K>0,\ |Ky|\ll1
&\Longrightarrow&
\mbox{Riccati/von Foerster-type growth},\\[1mm]
K>0
&\Longrightarrow&
\mbox{compressed-exponential growth}.
\end{array}
\label{eq:growth_hierarchy}
\end{equation}

The importance of the full TZ equation is therefore not that it replaces these
classical population models by an unrelated growth law. Rather, it embeds them within
a common nonlinear rate-feedback framework. Malthusian growth corresponds to
vanishing feedback; weak negative feedback produces the logistic limit and, more
generally, stretched-exponential growth; weak positive feedback produces the
Riccati/von Foerster limit and, more generally, compressed-exponential growth.
Retaining the full exponential feedback instead of truncating it allows the
instantaneous productivity to vary continuously with the state of the population and
thereby describes nonlinear growth regimes beyond the classical low-order
approximations.

This hierarchy is particularly useful for the human--android problem studied here.
Conventional demographic or manufacturing growth, resource-limited growth, and
self-amplifying technological production can all be interpreted as different regimes
of the same underlying state-dependent-rate principle.

\bibliographystyle{elsarticle-num}
\bibliography{references}

\end{document}